\documentclass[conference]{IEEEtran}
\IEEEoverridecommandlockouts

\usepackage{graphicx,xcolor}
\usepackage{subfig}
\usepackage{amsmath}

\usepackage{amsthm}
\usepackage{mathtools}
\usepackage{booktabs}
\usepackage{lipsum}
\usepackage{microtype}
\usepackage{url}
\usepackage[vlined,ruled,boxed,linesnumbered,commentsnumbered]{algorithm2e}
\usepackage{tikz}
\usepackage[final]{pdfpages}
\usepackage{extarrows} 
\usepackage{booktabs}
\usepackage[noadjust]{cite}
\usepackage{hyperref}
\usepackage{xspace}
\usepackage{array,multirow,multicol}

\usepackage{listings}
\usepackage{xcolor}
\usepackage{textcomp}
\usepackage{dblfloatfix}%
\usepackage{placeins}

\newif\ifislongreport
\islongreporttrue

\definecolor{commentgreen}{RGB}{2,112,10}
\definecolor{eminence}{RGB}{108,48,130}
\definecolor{weborange}{RGB}{255,165,0}
\definecolor{frenchplum}{RGB}{129,20,83}

\lstdefinelanguage{CAL}{
    classoffset=0,
	keywordstyle=\bfseries,
	morekeywords={namespace, external, action, network, entities, structure, actor,bitand,else,end,repeat,function, import, package, unit, procedure,priority,schedule,guard,var,do, if,lshift,not,then,size,type,for,foreach,while,fsm,begin,{==>}},
	keywordstyle=[2]\bfseries,
	keywords=[2]{int,uint,List,bool},
	classoffset=1, 
	classoffset=0,%
	basicstyle=\scriptsize\ttfamily,
	numberstyle=\tiny\ttfamily,  
	morecomment=[s]{/*}{*/},%
	morecomment=[l]{//},%
	morestring=[b]", 
	morestring=[b]', 
	captionpos=b,
	 emphstyle={\color{blue}},
	float=htpb, 
	xleftmargin=.001\textwidth, 
}

\newcommand{\Cal}{{\sc Cal}\xspace}

\begin{document}

\title{StreamBlocks: A compiler for heterogeneous dataflow computing (Technical Report)}

\author{\IEEEauthorblockN{
    Endri Bezati\textsuperscript{\textsection}, 
    Mahyar Emami\textsuperscript{\textsection}
}
	\IEEEauthorblockA{EPFL
	\\ Lausanne, Switzerland}
	\and
	\IEEEauthorblockN{Jorn W. Janneck}
	\IEEEauthorblockA{Lund University
	\\ Lund, Sweden}
	\and
	\IEEEauthorblockN{James R. Larus}
	\IEEEauthorblockA{EPFL
	\\ Lausanne, Switzerland}
}

\maketitle

\begingroup\renewcommand\thefootnote{\textsection}
\footnotetext{Equally contributing authors}
\endgroup

\begin{abstract}
To increase performance and efficiency, systems use FPGAs as reconfigurable accelerators. A key challenge in designing these systems is partitioning computation between processors and an FPGA. An appropriate division of labor may be difficult to predict in advance and require experiments and measurements. When an investigation requires rewriting part of the system in a new language or with a new programming model, its high cost can retard the study of different configurations. A single-language system with an appropriate programming model and compiler that targets both platforms simplifies this exploration to a simple recompile with new compiler directives.

This work introduces StreamBlocks, an open-source compiler and runtime that uses the \Cal dataflow programming language to partition computations across heterogeneous (CPU/accelerator) platforms. Because of the dataflow model's semantics and the \Cal language, StreamBlocks can exploit both thread parallelism in multi-core CPUs and the inherent parallelism of FPGAs. StreamBlocks supports exploring the design space with a profile-guided tool that helps identify the best hardware-software partitions.
\end{abstract}
 
\begin{IEEEkeywords}
actor machine, \Cal, dataflow, FPGA, HLS
\end{IEEEkeywords}

\section{Introduction}
\label{sec:introduction}

The slowdown in performance gain of general-purpose processors has increased interest in using FPGAs as reconfigurable accelerators, particularly for cloud computing~\cite{Putnam:2014, Caulfield:2016, Byma:2014, Weerasinghe:2015, Weerasinghe:2016}.

A key challenge in designing these accelerators is partitioning computation between processors and an  FPGA.
An appropriate division of labor may be difficult to predict and only be revealed by experiments and measurements.
When such an experiment requires rewriting part of a system in a new language, or with a new programming model, to run on the other platform, the high cost of exploration may retard studies of different configurations and limit the evaluation.
A possible solution is a single-language system with an appropriately portable programming model and a compiler to generate code for both platforms.
In this case, exploring a new system configuration just entails recompiling with new compiler directives.

A fundamental aspect is the programming model, which must allow parts of a program to run on either a CPU or an FPGA and interact transparently, regardless of where they execute.

The widely used \emph{accelerator model} fails in two aspects.
The different programming models for the CPU and FPGA make retargeting and design-space exploration costly and asymmetric.
It requires rewriting a function in a new language or dialect to reimplement it for hardware.
Also, it only supports a limited control transfer and communication model in which the CPU passes tasks to the accelerator.
A component running on an FPGA cannot use the CPU as a coprocessor.
However, industry flagship tools such as Xilinx SDAccel/Vitis and Intel OpenCL for FPGAs support this model for FPGA accelerators and MPSoCs.

Hardware could support more general models, such as concurrently processing multiple requests with pipelining or even out-of-order execution. However, these models integrate poorly with sequential programming languages such as C/C++ that natively support only sequential function calls.

One approach adopted by High-Level Synthesis (HLS) tools~\cite{webm_mentor, matlab_hdlcoder, 5g_xilinx} to support other models is to introduce streaming or dataflow programming through a library.
The hardware synthesized from C functions (\emph{C-HLS}) runs concurrently with the software components and communicates through FIFO channels.
The sequential semantics of C/C++, however, imposes limitations on the streaming functions~\cite{ug902}: 
(i) streams can only be single-producer, single-consumer~\cite{legup}, 
(ii) streaming functions cannot be bypassed, 
(iii) no backward links (feedback) between functions, and
(iv) conditional reading of a stream may prevent concurrent execution.
In other words, some patterns of execution that are natural for hardware are not expressible in a C function.

A workaround is to independently synthesize multiple streaming C-HLS functions with an HLS tool and manually connect the generated RTL modules with queues.
Each design requires a custom script to produce its streaming network,
and this script must evolve in step with the design.
Moreover, because the network of streaming functions may not have an explicit enter (call) or exit (return) point, 
unlike a C-HLS function, code must be rewritten to handle end-of-stream marks.

This work introduces an alternative approach based on StreamBlocks, a compiler suite for heterogeneous dataflow programs.
It enables the seamless hardware or software execution of streaming functions written in a \textit{single} source language.

StreamBlocks starts with a dynamic dataflow programming language (\Cal~\cite{EJ03}).
A dataflow program is a directed graph (cyclic or acyclic) whose nodes are operators called \emph{actors} and edges are streams.
A dataflow program specifies a partial order of computation, 
in which sequencing constraints arise only from data dependencies.
As a result, actors can execute concurrently.

A dataflow program can be compiled to run on a processor, FPGA, or a combination of the two.
In addition, dataflow semantics do not constrain an FPGA to operate only as a simple call-respond accelerator.
It allows the FPGA to operate as a streaming coprocessor that executes concurrently with the processor and possibly invokes operations implemented in software.
This generality allows the direct migration of part of a computation from software to hardware without rewriting the (dataflow) program, a essential step in developing, evaluating, or evolving a heterogeneous system.

However, the design space of possible partitions of functionality between the two platforms can be large and complex.
StreamBlocks includes a profile-guided tool to help identify the best-performing ones.

The contributions of this work are:
\begin{itemize}
    \item An open-source dataflow compiler for the \Cal programming language that targets heterogeneous platforms.
    \item An automatic partitioning methodology for placing computations on heterogeneous platforms without rewriting code.
    \item A tool that uses this flexibility to find high-performing hardware-software partitions of a system.
    \item Experimental results demonstrating that our compiler can efficiently use the resources of a heterogeneous platform. 
\end{itemize}

The paper is organized as follows.
Section~\ref{sec:dataflow} provides an overview of dataflow programming and the actor machine. Section~\ref{sec:hetero} presents the StreamBlocks compiler, heterogeneous code generation, and partitioning. Section~\ref{sec:related_work} compares StreamBlocks with other systems. Section~\ref{sec:evaluation} evaluates the compiler and partitioning. The paper concludes with Section~\ref{sec:conclusion}.

\section{Dataflow, actors, and actor machines}
\label{sec:dataflow}

\begin{figure}[h]
	\centering
	\includegraphics[width=\linewidth]{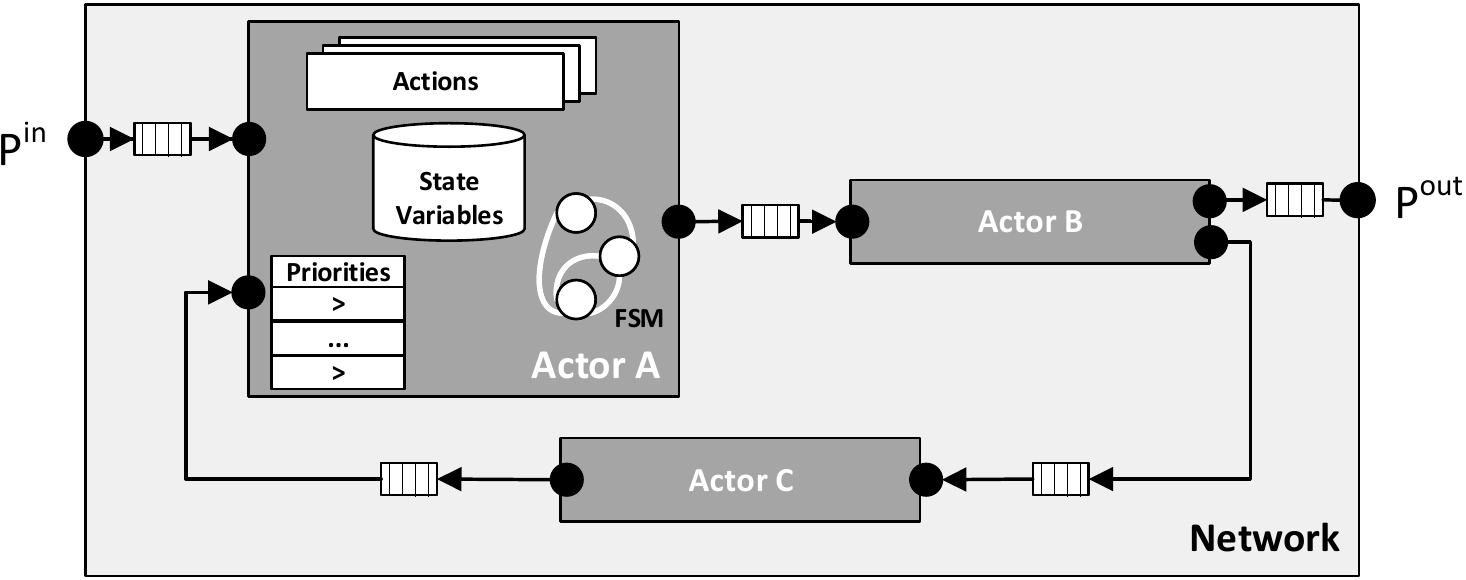}
	\caption{Structure of a \Cal dataflow program.}
	\label{fig:cal_model}
\end{figure}

Dataflow is a computation model used to express concurrent algorithms operating on streams of data.
It has a rich history with many variants (e.g., \cite{Dennis:1974,Kahn:1974,Lee:1995,Bilsen:1995}) often targeted at a specific application domain or designed to facilitate efficient analysis or implementation.
In our work, we use a general form of dataflow, in which a program is a \emph{network} of computational kernels, called \emph{actors}, that are connected via input and output \emph{ports} by directed, point-to-point connections or \emph{channels} (e.g., Fig.~\ref{fig:cal_model}).
Actors may have an internal state that is invisible to other actors.
They interact exclusively by sending packets of data called \emph{tokens} along the channels.
Channels are lossless and preserve the order in which tokens are sent.
They are also buffered (conceptually with unbounded capacity) so that the producer and consumer of a token need not synchronize during a transfer. A token may be consumed any time after it is produced.

Similar to the model introduced by Jack Dennis~\cite{Dennis:1974}, the actors considered in this paper execute in a sequence of atomic \emph{steps} or \emph{firings}.
During each step, an actor may (a) consume input token(s), (b) produce output token(s), and (c) modify its internal state.
Instead of being written as a sequential process (e.g., Kahn processes~\cite{Kahn:1974}), languages such as \Cal are structured around the steps an actor may perform.
In \Cal, an actor is a collection of \emph{actions}, each describing a step that the actor can perform, along with the \emph{conditions} under which the step should execute.\footnote{Processes can be translated into a \Cal-like description, cf. \cite{Cedersjo:2019}.}

\subsection{\Cal Actor Language}
\label{sec:cal}

The \Cal actor language~\cite{EJ03} is a programming language for expressing actors with firing, the subject of this paper. It permits the description of actors for a wide range of dataflow models, ranging from Kahn process networks~\cite{Kahn:1974}, dataflow process networks~\cite{Lee:1995}, and synchronous and cyclo-static dataflow~\cite{Lee:1987,Bilsen:1995,Buck:1993}, among others. Parts of the language have been standardized by the ISO/IEC committees \emph{RVC-CAL}~\cite{iso230014} as one of the languages used to write the reference implementations of MPEG video decoders.

\begin{lstlisting}[language=CAL,belowcaptionskip=-15pt, caption=A simple network of two actors in \Cal.,label={cal:filter}]
namespace example:

  external function rand(int x) --> int end

  actor Source() int IN ==> int OUT:
    int x := 0;
    action ==> OUT:[rand(x)] 
    guard
      x < 4096
    do 
      x := x + 1; 
    end
  end

  actor Filter(int param) int IN ==> int OUT:

    function pred(int param, int value) --> bool :
      if (param > value) then true else false end
    end

    t0: action IN:[t] ==> OUT:[t] guard pred(param, t) end
    t1: action IN:[t] ==> end

    priority
      t0 > t1;
    end
  end

  actor Sink() int IN ==> :
    action IN:[x] ==> do println(x) end 
  end
 
 network TopFilter(int param) ==>:
    entities
      source = Source();
      filter = Filter(param=param);
      sink   = Sink();
    structure
      source.OUT --> filter.IN;
      filter.OUT --> sink.OUT;
  end
end
\end{lstlisting}

Listing~\ref{cal:filter} contains a \Cal dataflow program composed of four entities: three actors and a network. Actor \texttt{Source} has a single output port \texttt{OUT}. Its single action describes the only step it takes, incrementing a state variable \texttt{x} up to 4096 and producing a token to be sent to its output port. The token contains value from invoking the external function (\texttt{rand}) on \texttt{x}. External functions provide access to libraries and legacy code in a dataflow program. 

The \texttt{Filter} actor's parameter, \texttt{param}, is a number used by the local function \texttt{pred}. \texttt{Filter} uses an input port \texttt{IN} and an output port \texttt{OUT}. It comprises two actions. The first, labeled \texttt{t0}, includes a \emph{guard} condition, a logical expression that must be true (in addition to an input token being available) for the actor to be able to take the action---copying the input token to the output. The second action, labeled \texttt{t1}, also consumes an input token but has no guard and produces no output. A priority rule specifies that action \texttt{t0} has a higher priority than action \texttt{t1}. This means that whenever conditions for both are satisfied, action \texttt{t0} executes. In this way, action \texttt{t0} copies to the output all input tokens that satisfy its guard, while action \texttt{t1} ``swallows'' the other tokens. Finally, Actor \texttt{Sink} consumes a token from its input port \texttt{IN} and prints it to the console.

 The \texttt{TopFilter} network connects the three actor instances. The dataflow network in \texttt{TopFilter} has no input and output ports. It has three entities \texttt{source}, \texttt{filter}, and \texttt{sink}, which are instantiations of actors \texttt{Source}, \texttt{Filter}, and \texttt{Sink}, respectively. The \texttt{filter} instance is instantiated with a parameter received from the network.

The \texttt{TopFilter} network connects the \texttt{IN} input port of the network to the input \texttt{IN} of the \texttt{rand} instance. The output port \texttt{OUT} of instance \texttt{source} is connected to the input port \texttt{IN} of \texttt{filter} instance. The \texttt{filter} output port \texttt{OUT} is connected to the input port \texttt{IN} of the instance \texttt{sink}. This paper will use this \Cal example to explain our work.

\subsection{Actor Machines}
\label{sec:am}
\label{subsec:am}
Executing an actor such as \texttt{Filter} alternates between two phases: (1) selecting the next action to execute and (2) executing it. In general, the choice of the next action can depend on many conditions (i.e., availability of input tokens, satisfiability of action guards, and availability of space in the output buffers) and might be ordered by priorities. At each point in time, zero, one, or more actions can be executed. If zero, action selection waits for an external event (arrival of input or freeing of output buffer space) to make an action executable. If more than one action can be executed, any can be chosen.

The \emph{actor machine} (AM)~\cite{Janneck:2011} is an abstract machine model for actors that codifies the action selection process in a state machine called its \emph{controller}. While an action's execution is a single atomic step of an actor, it usually requires several microsteps of an AM to evaluate and test (and possibly re-evaluate and re-test) conditions until it can finally select and execute an action.

\begin{figure}[!h]
	\centering
	\includegraphics[width=\linewidth]{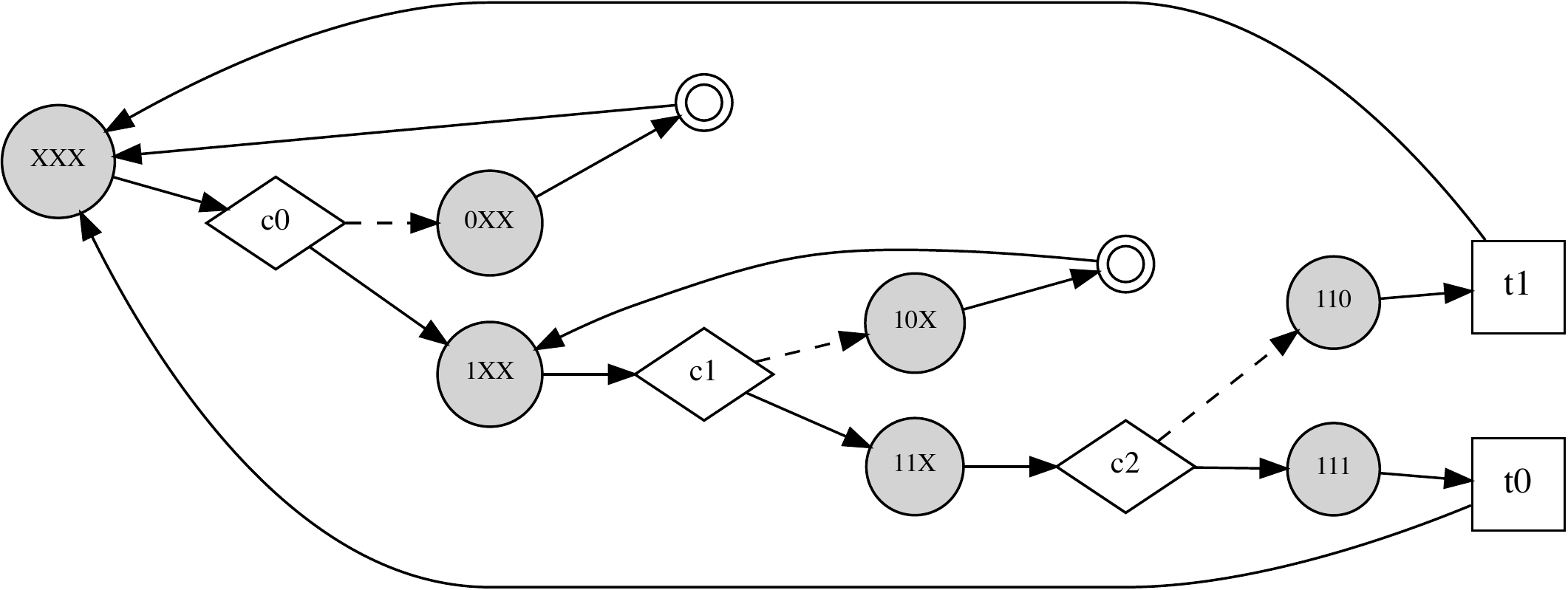}
	\caption{Controller state machine for the \texttt{Filter} actor in Listing~\ref{cal:filter}. The gray circle, 
	diamond, square, and ring respectively represent state, \texttt{TEST}, \texttt{EXEC}, and \texttt{WAIT}.}
	\label{fig:filter_actor}
\end{figure}

Fig.~\ref{fig:filter_actor} shows a controller for the \texttt{Filter} actor from Listing~\ref{cal:filter}. Its execution depends on three conditions: $c_0$, the availability of a token on the input port, $c_1$, the availability of space on the output port, and $c_2$, the guard of the first action. In the figure, the shaded circles represent the \emph{controller states}, while the diamonds, boxes, and rings represent the three kinds of controller \emph{instructions} that transition the controller from one state to another:
\begin{enumerate}
	\item \texttt{TEST}: tests a condition (either presence of input, availability of output space, or a guard condition) and takes the controller to one of two states depending on whether the condition is true or false, 
	\item \texttt{EXEC}: executes an action and transitions to a new controller state,
	\item \texttt{WAIT}: transitions to a new controller state.
\end{enumerate}
Only an \texttt{EXEC} microstep changes the program-defined state of an actor machine (i.e., the internal state variable assignments and the streams connected to its ports).

Every state of an AM controller corresponds to a specific configuration of knowledge about the conditions, each of which may be true, false, or unknown (represented by $1$, $0$, and $X$, respectively).  In the example, each controller state is labeled with a triple of values from $\{0, 1, X\}$, corresponding to what is known about $c_0$, $c_1$, and $c_2$ at that point in the controller's execution. For example, the initial controller state, when nothing is known, is labeled $XXX$. Once the first condition, $c_0$ is tested and it is known whether there is an input token, the controller transitions to either $0XX$ (no input token) or $1XX$ (input token available). If in the process of checking conditions, the controller ends up in a state in which it can perform an \texttt{EXEC} instruction, it has successfully selected an action. The \texttt{EXEC} instruction then executes the action, and the selection process begins anew.

If the controller ends up in a state such as $0XX$, nothing can be done until an input token arrives. When something outside of the AM changes, the AM condition needs to be re-tested. This means the AM controller needs to ``forget'' that there was no input token and return to the initial state $XXX$ to re-test. This is the role of the \texttt{WAIT} instruction, which ``throws away'' knowledge about transient conditions (i.e., conditions that can change through things happening outside of the AM, such as tokens arriving or output space becoming available), so they can be re-assessed. A \texttt{WAIT} instruction is typically used when the controller comes to a point at which it cannot execute an action or test additional conditions that would enable it to perform an action. In other words, the controller is stuck until some external occurrence changes one of its transient conditions.

The AM in Fig.~\ref{fig:filter_actor} contains exactly one instruction in each state (a so-called single instruction AM, or SIAM). In general, AMs that allow a choice of multiple instructions to proceed from a controller state (multiple instruction Actor Machine, or MIAM) makes the controller non-deterministic. In this paper, the actor machines of a CAL actor use only a single instruction (SIAM) for each state to reduce the number of tests needed at runtime. A detailed discussion of the relationship between single instruction and multiple instruction machines can be found elsewhere~\cite{Cedersjo:2012}. All that matters for this paper is that any MIAM can be reduced to a SIAM that maintains the actor's original semantics while possibly reducing the behaviors it may exhibit.

\subsection{AM Network and Idleness}
\label{subsec:am_net}

AMs can be connected to create a \textit{network} of AMs.
The \texttt{TopFilter} actor in Listing~\ref{cal:filter} is an AM network expressed in CAL, with 
two actors \texttt{rand} and \texttt{filter}.
As apparent in Listing~\ref{cal:filter}, so long as the \texttt{Filter} actor
has input, it continues to take actions. In fact, this actor never terminates but simply waits when its input
stream is empty. It resumes taking actions when additional tokens are fed into its input. If
an actor stops working due to an absence of external events, we say the actor became \textit{idle}.

With a single AM, the controller state machine can check
for idleness. An AM is idle if its controller state machine can only perform sequence of \texttt{TEST} steps ending with a \texttt{WAIT} in absence of external events.
However, reasoning about idleness in a \emph{network of AMs} is more subtle, especially 
in presence of cycles in the network and unpredictable token production/consumption,
a characteristic of AMs.

Idleness is an important property because if a hardware AM network can autonomously detect its idleness, 
then software need not continuously poll to see if hardware has completed an operation.
Since, in general, a hardware network can produce a (statically) unknown number of output tokens, without a method to detect when the hardware is finished (idle), software must continually poll for results or
end-of-streams be explicitly baked into the code by programmers. 
The StreamBlocks compiler implements a run-time mechanism 
to efficiently detect idleness on hardware without software intervention.


\begin{figure*}[t]
	\centering
	\includegraphics[width=0.9\linewidth]{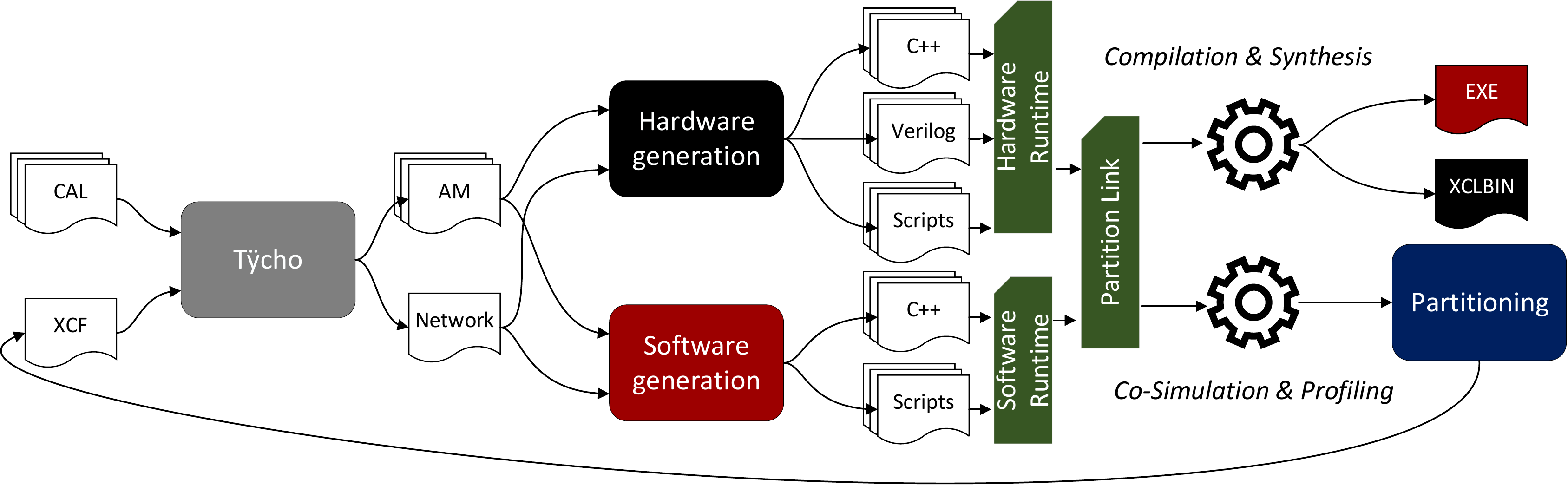}
	\caption{StreamBlocks heterogeneous compilation flow. The compilation and synthesis results in a multi-threaded executable that asynchronously launches an OpenCL RTL kernel (XCLBIN). The XCF configuration file can either be specified by a user or generated automatically by the partitioning tool..}
	\label{fig:streamblocks}
\end{figure*}

\section{StreamBlocks}
\label{sec:hetero}

StreamBlocks is a suite of tools for designing, compiling, and optimizing streaming dataflow applications. 
The framework consists of three parts: \Cal compiler front-end, back-end code generators, and a partitioning tool. 
The \Cal front-end comes from T\"ycho~\cite{Cedersjo:2019} and supports a common representation based on the AM machine model on which to perform transformations and optimizations. 
Both imperative and
action-based kernels can be translated to this representation, making the same optimizations applicable
to different types of kernels and across language boundaries. 
StreamBlocks extended T\"ycho with a type system that supports \textit{Product} and \textit{Sum}
types similar to those in functional programming languages.

T\"ycho produces an AM representation, which is fed to the homogeneous/heterogeneous StreamBlocks back-ends
to apply platform-specific transformations (Section~\ref{sec:hetero:hw} and \ref{sec:hetero:sw}). 
Each platform supports a runtime responsible for actor scheduling. 
Hardware actors are scheduled in parallel whereas software actors are scheduled on cores following a
user-provided, actor-to-core mapping.
An XML configuration file specifies these actor-to-core or actor-to-FPGA mappings. 

The final aspect of StreamBlocks is partitioning (Section~\ref{sec:hetero:partitioning}).
Each runtime supplies methods for profiling a dataflow application on its platform (Section~\ref{sec:hetero:profiling}). Profiling uses either hardware cycle timers
for CPUs or cycle counts from a cycle-accurate RTL simulation. Also, StreamBlocks provides tools for profiling inter-actor communication on a heterogeneous platform. 
The StreamBlocks partitioning uses profiling information to map actor instances to the processing cores in a homogeneous platform or across processing cores and FPGA on a heterogeneous platform.

Finally, this paper focuses on software and hardware code generation for heterogeneous platforms that support the Xilinx OpenCL runtime, i.e., any Xilinx PCIe accelerator board or MPSoC.

\begin{lstlisting}[language=XML,caption=Configuration file for the \texttt{TopFilter} network.,label={xcf:filter}]
<?xml version="1.0" encoding="UTF-8" standalone="yes"?>
<configuration>
 <network id="example.TopFilter"/>
 <partitioning>
   <partition id="0" pe="x86_64" 
                     code-generator="sw">
     <instance id="sink"/>
   </partition>
   <partition id="1" pe="xczu7ev-ffvc1156-2-e" 
                     code-generator="hw">
     <instance id="source"/>
     <instance id="filter"/>
   </partition>
 </partitioning>
 <code-generators>
   <code-generator id="sw" platform="multicore"/>
   <code-generator id="hw" platform="vivado-hls"/>
 </code-generators>
 <connections>
   <fifo-connection source="source" source-port="OUT" 
                    target="filter" target-port="IN" 
                    size="1"/>
   <fifo-connection source="filter" source-port="OUT" 
                    target="sink" target-port="IN" 
                    size="4096"/>
 </connections>
</configuration>
\end{lstlisting}

\subsection{Configuration File}
\label{sec:hetero:xcf}

A developer can use source-code annotations to force the placement of \Cal actors on either hardware or software. It is also possible to partition actors between hardware and software using a StreamBlocks XML Configuration File (XCF). For example, Listing~\ref{xcf:filter} shows a mapping of the \texttt{TopFilter} network from Listing~\ref{cal:filter}. \texttt{Source} and \texttt{Filter} execute on the FPGA and \texttt{Sink} on the CPU. 
Most actors can run on either platform, but some cannot be placed on hardware, for example, an actor that reads a file.

The configuration file specifies the \Cal network partitioning, code generators, and connections. 
Each partition has an name (id), the id of the code generator to use, and the names of the \Cal instances it contains.
A code generator can be configured with platform-specific settings (i.e., desired clock frequency, maximum BRAM size, \dots). 
If these are not specified, the compiler uses a default value.
There are two types of connections: FIFO connections and external memory connections. 
If the depth of a FIFO connection is not declared, the code generator is free to choose any value. 
External memory connections are used to hold large actor state variables, which do not fit in BRAM, in a memory 
bank on the FPGA PCIe accelerator board.

\subsection{Hardware Code Generation}
\label{sec:hetero:hw}

The StreamBlocks compiler generates a C++ class for each Hardware AM (HAM). Listing~\ref{cpp:filter_class} contains 
a class definition for actor \texttt{Filter} in Listing~\ref{cal:filter}. The top function for HLS is \texttt{operator()}, which implements the AM controller state machine. 
The \texttt{transition} and \texttt{condition} functions implement the \texttt{EXEC} and \texttt{TEST} steps. The \texttt{scope} function sets variable assignments for actions and is invoked before the \texttt{transition} or \texttt{condition} functions.

We use  Vivado HLS's streaming library, \texttt{hls::stream} class, to implement the input and output ports for AMs.
To execute an action, a controller state machine checks if (i) input tokens are available, (ii) there is space
for the output, (iii) and guard conditions are satisfied.  \texttt{hls::stream} does not offer appropriate functionality since 
its API does not allow reading the size of a stream or peeking at values without consuming them.
To avoid these limitations, we created a custom Verilog First-Word Fall-Through (FWFT) FIFO whose outputs are its size, 
count, and head queue element and that is compatible with the \texttt{hls::stream} RTL input/output interface. To use these output values, 
a unique \texttt{IO} structure for each AM is generated and the count, size and queue are passed by value (\texttt{io} argument of the top HLS function) as shown in Listing~\ref{cpp:filter_class}.

The HLS implementation of a controller state machine follows its definition in Section~\ref{subsec:am}.
At each invocation of \texttt{operator()}, the actor can take multiple steps or instructions. Its state is recorded in the variable \texttt{program\_counter} to allow execution to continue between invocations.

\begin{lstlisting}[language=C++, caption=C++ class and top HLS function for the \texttt{Filter} AM (Vivado/Vitis HLS).,label={cpp:filter_class}]
struct IO{
  int32_t IN_peek;
  int32_t IN_count;
  int32_t OUT_size;
  int32_t OUT_count;
};

class class_filter {
private:
  // -- State Variables
  bool pred_1(int32_t l_param, int32_t l_value);
  int32_t a_param, a_t__1, program_counter;
  
  // -- Scopes
  void scope_1(IO io);

  // -- Conditions
  bool condition_0(hls::stream< int32_t > &IN, IO io);
  bool condition_1(hls::stream< int32_t > &OUT, IO io);
  bool condition_2();

  // -- Transitions
  void transition_0(hls::stream< int32_t > &IN, 
                    hls::stream< int32_t > &OUT);
  void transition_1(hls::stream< int32_t > &IN);

public:
    // -- For static instatiation
    class_filter(int32_t param){ a_param = param; }
    // -- Controller
    int32_t operator()(hls::stream< int32_t > &IN, 
                       hls::stream< int32_t > &OUT, IO io);
};

// -----------------------------------------------
// -- HLS Top Function, instantiated with param=10
int32_t filter(hls::stream< int32_t > &IN, 
           hls::stream< int32_t > &OUT, 
           IO io){
    static class_filter i_filter(10);

    return i_filter(IN, OUT, io);
}
\end{lstlisting}

Fig.~\ref{fig:filter_actor} depicts the example's controller state machine. 
This controller has two \textit{entry} points, state $XXX$ and $1XX$, and 4 \textit{exit} points, when the controller 
reaches the two wait states and after executing $t_0$ or $t_1$.
Suppose that on invocation, the controller is in state XXX and all conditions $c_0$ to $c_2$ are true. Then, the controller performs \texttt{TEST} steps for $c_0$, $c_1$, and $c_2$, changing state accordingly to 1XX, 11X, 111. At state 111, it
performs the \texttt{EXEC} step $t_0$, transitions to state XXX, returns a code that indicates its last step was an \texttt{EXEC}, and exits.
The controller is acyclic; it does not execute the same step multiple times in a single invocation. Because the steps are bounded, the HLS scheduler (i.e., Vivado HLS) can schedule all \texttt{TEST} steps, $c_0$ to $c_2$, at the first clock cycle of invocation. To improve performance, the controller takes the maximum number of steps possible (without cycling) before it exits.

StreamBlocks automatically inserts some HLS directives (i.e., \texttt{pragma}) in appropriate places to optimize the hardware.
For instance, all scope, condition, and transition methods are inlined in the controller. Also, loops with fewer
than 64 iterations are unrolled to accelerate \texttt{repeat} expressions without incurring an unreasonable resource cost.
A developer can also directly annotate the source code with other directives. For example, \textit{@loop\_merge} can be used to merge input-read and output-write loops, and \textit{@external} can place an actor variable in an off-chip memory (e.g., DDR or HBM). In the latter case, the compiler produces appropriate HLS pragmas to add an AXI master interface to the actor.

FIFOs that cross the hardware/software boundary are connected to \textit{Input} and \textit{Output Stage} actors. 
These special actors transparently pass tokens between hardware and software without developer involvement (Section~\ref{sec:hetero:plink}).

Vivado HLS compiles each actor to RTL.
The corresponding RTL actor is instantiated in a Verilog
network along with its \textit{trigger} module. 
The trigger is a hardware scheduler that enables or disables its actor.
In hardware, ideally, all actors should execute asynchronously to maximize performance.
However, hardware should also detect when forward progress is no longer possible (idleness) and inform software that computation is complete and output data is available.

\begin{figure}[h]
	\centering
	\includegraphics[width=\linewidth]{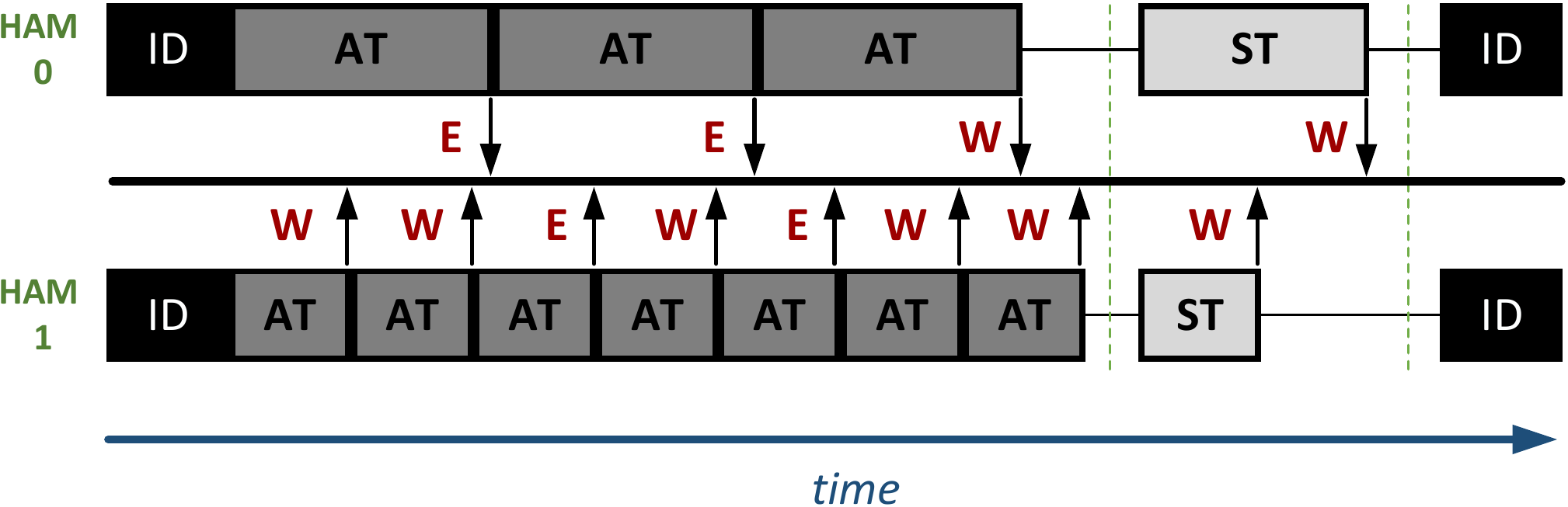}
	\caption{An example trace of the execution of two hypothetical HAMs.
	\texttt{ID, AT, ST} corresponds to idle, asynchronous trigger and synchronous triggering. \texttt{E, W} 
	corresponds to \texttt{EXEC}, \texttt{WAIT} return codes from a HAM.
	Vertical green lines represent synchronization points.}
	\label{fig:ham_triggering}
\end{figure}

We use the triggers to detect idleness.
Detecting idleness in an actor network requires synchronized coordination. 
The triggers continually monitor network state and eventually perform a synchronized idleness check.
Fig.~\ref{fig:ham_triggering} illustrates an example. \texttt{ID}, \texttt{AT}, and \texttt{ST} 
correspond to idle, asynchronous triggering, and synchronous triggering, respectively. In the normal state, \texttt{AT}, 
actors are repeatedly and asynchronously activated until all triggers in the network observe their actors are idle.
When that happens, the triggers synchronously execute \texttt{ST} to check idleness and disable the network if it is the case.
For conciseness, we omit details of our hardware idleness detection algorithm.

In the Verilog network, actors communicate through FWFT queues.
When the first token is written into a queue, 
it immediately appears on the output, thus it allows the token to be read on the next clock cycle. The FIFO's data width is set automatically from the width of the outgoing and incoming ports of the connected actors. If the outgoing and incoming ports' width differ, the compiler reports an error. The depth of all queues in the network is set to a compiler-defined value if the XCF configuration file does not specify it.

\subsection{Software Code Generation}
\label{sec:hetero:sw}

Similar to hardware code generation, the StreamBlocks compiler translates a software AM (SAM) to a C++ class that runs on a processor. 
Our multi-threaded runtime instantiates each actor class.
Conceptually, it seems appropriate to allocate a thread to each actor since they execute independently.
However, the relatively high cost of context switches leads to sub-optimal performance since an actor's controller may not run for long before it determines that the actor is still idle.
A better use of resources is to map several actors onto a thread and run them sequentially such that the cost of a context switch can be amortized over a larger computation.
However, if several of these actors could run concurrently, then the opportunity to improve performance by running in parallel is lost.
The actor-to-thread mapping is specified in the XCF file.
The threads are assigned to dedicated physical cores.
Each pinned thread has an independent, round-robin scheduling loop for its actors. 
If no thread mapping is specified in the XCF, a single thread runs all actors.

Unlike hardware execution, in which the controller state machine cannot have cycles in a single invocation,
to reduce scheduling overhead,
the software controller state machine performs as many steps as possible.
It may execute the same step multiple times, 
before yielding to the scheduler.
The compiler has a default maximum threshold for iterating in the controller, which can 
be overridden by command-line arguments. In the case that an actor reaches a \texttt{WAIT} instruction before the threshold is 
hit, it yields earlier.

FIFOs are implemented as ring buffers.
A structure attached to every AM output port holds global and local FIFO counters. The global counters are visible to all partition threads, but a local one is visible only to its thread.
Global and local counters do not change while the actors are running.
Updates are cached internally and synchronously applied after full iteration of the round-robin scheduling. This allows the FIFO ring buffers to be lock-less, as threads only see newly written/read-and-freed tokens after this step.

\begin{figure}[h]
	\centering
	\includegraphics[width=0.7\linewidth]{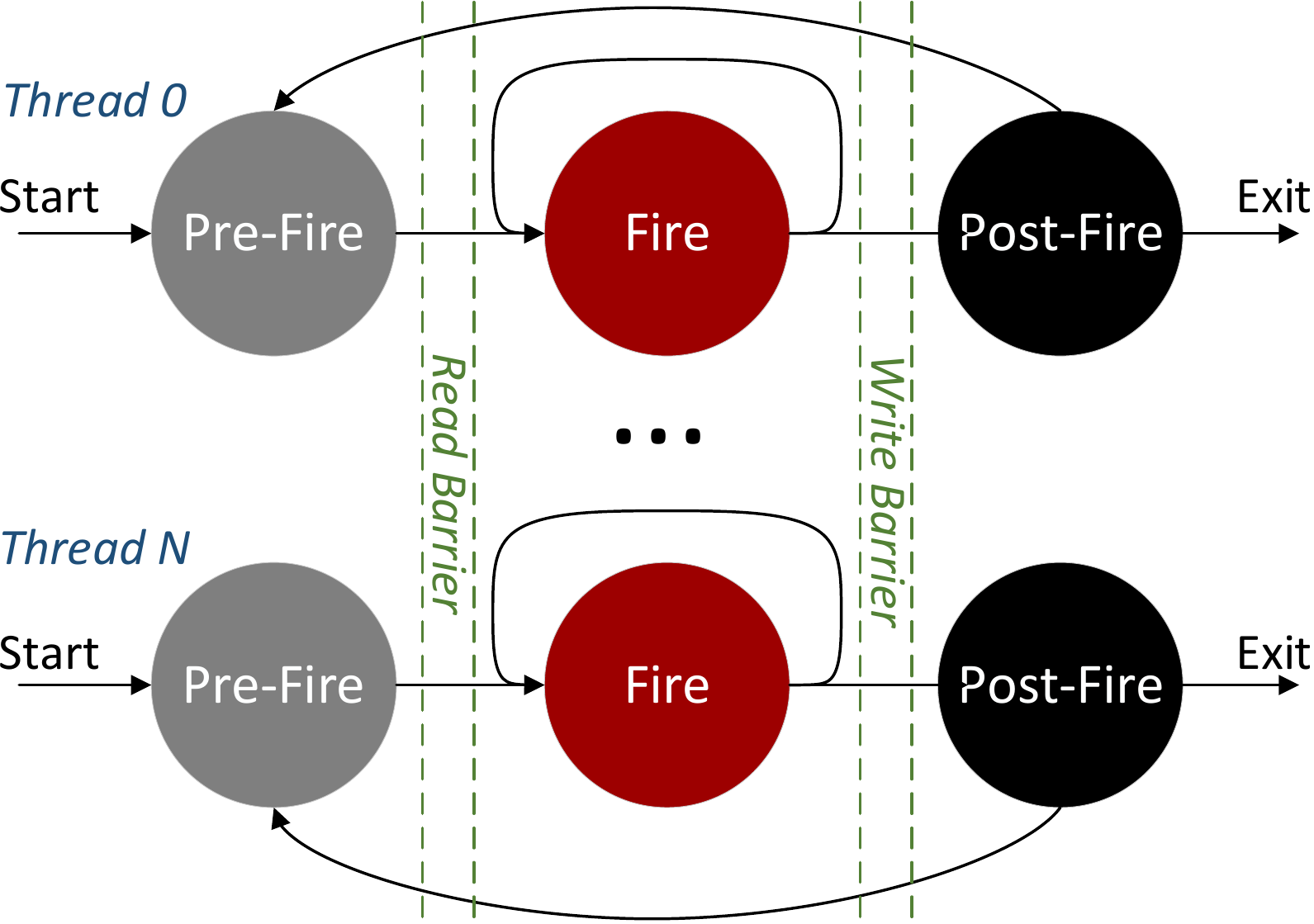}
	\caption{Pre-Fire, Fire, and Post-Fire steps for scheduling actors on a thread. \textit{Pre-Fire} checks if a thread has sufficient data to execute. \textit{Fire} runs the actors in the partition with a round-robin schedule, and \textit{Post-Fire} checks if the thread should iterate, sleep or terminate.}
	\label{fig:multicore_threading}
\end{figure}

Fig.~\ref{fig:multicore_threading} shows the three steps\footnote{Pre-Fire, Fire, and Post-Fire~\cite{Eker_Janneck:2003}.} of executing a thread in the software runtime. 
\textit{Pre-fire} determines how much data can be read and written and stores the local FIFO counters' values in its thread. 
\textit{Fire} runs all SAMs in the partition in a round-robin scheduling iteration and makes visible all FIFO writes from the SAMs that fired during the iteration. \textit{Post-fire} checks if a SAM executed in the \textit{Fire} step, so other threads should wake up. If no SAMs executed in the \textit{Fire} step, then the thread sleeps. If all other threads are asleep and all have had a ``quiescent'' round in which no tokens are produced or consumed, the threads terminate.

The software runtime provides the threading mechanism and FIFO implementation. In addition, it supports system operations---such as file operations, image/video visualization, and other functionality---and links with library and legacy functions.

\subsection{Input/Output Stages and Partition Link}
\label{sec:hetero:plink}

\begin{figure}[h]
	\centering
	\includegraphics[width=0.9\linewidth]{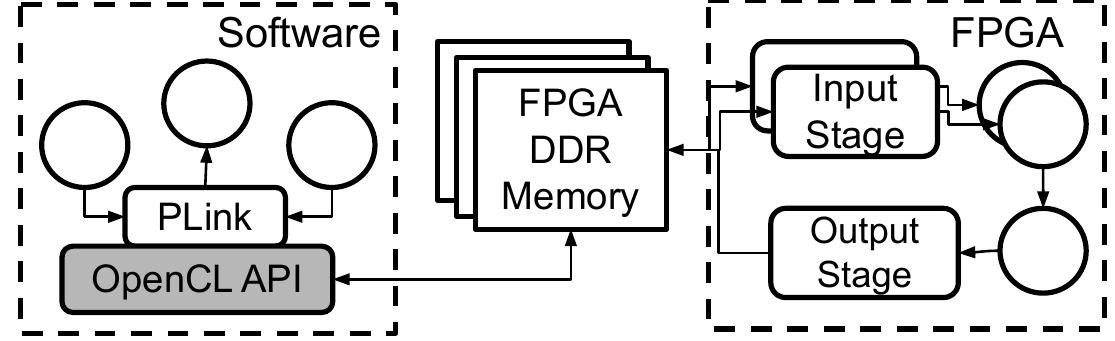}
	\caption{The hardware/software interface. Circles represent application actors.
	Arrows show communication paths. Input stage, output stage, and PLink are special actors
	that are instantiated depending on hardware/software actor placement.}
	\label{fig:arch}
\end{figure}

To enable actors to run on hardware or software, the compiler generates appropriate hardware/software interfaces. Fig.~\ref{fig:arch} depicts the interface architecture.

PLink (Partition Link) is a special actor that (1) transfers software FIFOs to FPGA DDR memory, 
(2) starts the execution of hardware, and (3) receives data from the FPGA memory when the FPGA network becomes idle.
A PLink treats the hardware network like a coprocessor and is its software controller.
It invokes the coprocessor so long as the software continues to provide input values.
Underneath, PLink takes advantage of OpenCL's API~\cite{OpenCL}, basically treating the hardware coprocessor as a \emph{kernel}.

On the FPGA partition, for every incoming communication channel from software, we instantiate an input stage actor.
These actors read data from the FPGA DDR memory in bursts which are then streamed to a BRAM FIFO connected to 
consumer actors.
The output stage actor in Fig.~\ref{fig:arch} carries out a similar task and reads an FPGA FIFO corresponding to 
an outgoing communication paths and writes its values to DDR memory for the PLink to read. 
The input stage, output stage, and hardware actors constitute the \emph{dynamic region} in 
Xilinx terminology. The dynamic region has multiple AXI memory ports with a single AXI slave port. The 
AXI slave port receives a start command from the PLink through OpenCL APIs (\texttt{clEnqueueTask}) and 
notifies it when the hardware execution becomes idle which is captured by the OpenCL runtime.
The dynamic region is automatically wrapped by a static \emph{shell} which connects
all the AXI ports to appropriate blocks through the RTL kernel 
compilation flow of Vitis~\cite{ug1393}. Bear in mind that the dynamic region changes by the hardware partition.

Additionally, when an FPGA actor requires DDR memory for object storage (e.g., an actor that uses a large array
as an internal variable), the compiler instantiates appropriate memory interfaces.
The PLink allocates the DDR memory on startup and passes the pointer to the hardware actor. 
Fig.~\ref{fig:arch} omits these details for clarity.

The PLink takes advantage of OpenCL events and an out-of-order command queue for nonblocking operations.
All buffer transfers through  \texttt{clEnqueueWrite} and \texttt{clEnqueueRead} and the starting of
the kernel through \texttt{clEnqueueTask} have corresponding events. Consequently, the PLink never blocks 
the software thread it is scheduled on so that other actors can perform useful work.

\subsection{Testing \& Profiling}
\label{sec:hetero:profiling}

The compiler provides a co-simulation solution that runs software actors on the software runtime and instantiates hardware actors as cycle-accurate SystemC models.
A developer can test their heterogeneous application by simply specifying a compiler switch.
We rely on Verilator to translate Verilog to SystemC.
PLink transparently establishes the interface between the software and SystemC simulation domains.
We do not yet model off-chip memory accesses, but will implement this feature in the future.

Co-simulation is also important for profiling hardware actors, which provides the basis for automatic hardware-software partitioning.
The triggers in the SystemC network can optionally profile individual hardware actors. A trigger logs the minimum, maximum, and average clock cycles that an actor spends executing actions.

Likewise, the software runtime can profile similar information.
The runtime uses the \texttt{rdtscp} time counter for x86 code, whereas ARM code uses the virtual \texttt{cntvct} counter scaled by clock frequency.

\subsection{Heterogeneous Partitioning}
\label{sec:hetero:partitioning}

Finding an appropriate balance between hardware and software execution is a design space exploration problem part of 
hardware-software partitioning or hardware-software co-design. 
StreamBlocks facilitates this exploration process since an actor written in \Cal can run in hardware or software with no code changes. 
In general, the design space is very large and raises several practical questions: 
(1) Which mapping of actors to software and FPGA produces the best performance? 
(2) Is it practical to find such partition(s) automatically?

Hardware/software partitioning can be modeled as a graph partitioning problem with a cost function.
In this paper, we focus on execution time as the single metric to optimize, but a practical design must also balance 
resource usage, power, and other considerations.
The cost function used in design space exploration should estimate the execution time for a given partitioning. Performance prediction 
is difficult for a variety of reasons: actors in \Cal exhibit dynamic (sometimes even non-deterministic) behavior,
the multi-threaded runtime can behave unpredictably, software-hardware crossings are difficult to model, 
and actors' behavior is data-dependent because of dynamic data dependencies in and among actors.

To estimate multi-threaded software-hardware performance, we used a mixed integer linear programming (MILP) model.
A model's structure reflects the corresponding actor network graph.
It is parameterized by profiling data (Section~\ref{sec:hetero:profiling}) and measurements of software FIFO throughput and OpenCL host-to-device and device-to-host transfers.

\ifislongreport
We briefly present our model here, for more details refer to Section~\ref{sec:appendix}.
\else
We briefly present our model without going into details.
\fi
Given a set of $n$ threads $P_{thread} = \{p_1, p_2, ...,p_n\}$, an FPGA partition ${accel}$,
and a set of actors $A$, we define the decision variables $d^{a}_{p}$ with the following constraints:

\begin{gather*}
    \forall a \in A, \forall p \in P_{thread}\cup \{{accel}\} : d^{a}_{p} \in \{0, 1\} \\
    \forall a \in A : \sum_{p \in P_{thread}\cup \{{accel}\}} d^{a}_{p} = 1
\end{gather*}
The first constraint states that the decision variables are boolean and the second ensures that
each actor $a$ maps to exactly one partition $p$.

The execution time of an actor $a$ on partition $p$ is given by $exec(a, p)$, which is obtained through software or
hardware profiling. The time spent on a thread partition is the sum of all actor execution times since actors run sequentially on a thread:
\begin{equation}
    \forall p \in P_{thread}: T_p = \sum_{a \in A} d^{a}_{p} \times exec(a, p)
\end{equation}

We delegate the performance modeling of a hardware actor to the PLink actor described in 
Section~\ref{sec:hetero:plink}. We assume that all PLink operations are asynchronous and, without loss of generality, we assume PLink is always scheduled by $p_1$.
\begin{equation}
\begin{split}
    T_{plink} & = max(\{d^{a}_{accel} \times exec(a, accel)\}) \\
              & + T^{read}_{plink} + T^{write}_{plink} 
\end{split}
\end{equation}

The first term models hardware execution time as the maximum of actor execution times since actors can run concurrently 
on the FPGA. The other two terms model OpenCL read and write transfers, which depend on the number of tokens and the OpenCL cost information.
Note that although PLink's operations are asynchronous, because there a logical dependence between write, execution
and read, their total time is their sum. 

Similar to the hardware performance, we model the multi-threaded execution time as the maximum of individual thread execution times:
\begin{equation} \label{eq1}
\begin{split}
    T_{exec} & = max(\{T_p : p \in P_{thread}\} \cup \{T_{plink}\})  \\
             & + T_{intra} + T_{inter}
\end{split}
\end{equation}

Observe that $T_{plink}$ is not added to $T_{p1}$ since it operates asynchronously, however, 
if the hardware is slower than all software threads, then execution time is determined by hardware,
 hence the $max$ operation.
The $max$ term models pure actor execution time with no regard for communication cost between actors.
Actors that communicate heavily can exploit fast caches if they are assigned to the same thread. 
Actors on different threads will communicate through the shared (e.g., L3) cache or main memory.
$T_{intra}$ and $T_{inter}$ model the latency of intra- and inter-core communication.
The latter two terms help ensure that actors that communicate heavily reside on the same thread.

The terms $T_{intra}$ and $T_{inter}$ are MILP formulas that depend on the software FIFO bandwidth 
and the number of tokens that are transferred on each FIFO. Our software runtime provides
this profiling information.

Using a MILP formulation enables us to incorporate other constraints into the search.
For instance, we can constraint the solutions to have fewer than $m$ FIFOs crossing the hardware-software boundary. 
This might be important when an FPGA platform has a limited
number of AXI master ports to off-chip memory.

\section{Related work}
\label{sec:related_work}
\Cal has been used in many studies that produce both software and hardware code. The earliest publication on hardware code generation is~\cite{Janneck:2009}. It used OpenDF and Forge HLS
from Xilinx. Orcc~\cite{Orcc:2013} is another framework for RVC-CAL that focused on video streaming applications.

Another tool, Xronos~\cite{Bezati:2013:ISPA}, used an open-source version of Forge HLS to translate \Cal to HDL, again using the Orcc
compiler. Compared to OpenDF, Xronos completely supports the RVC-CAL language for hardware synthesis. 

Exelixi~\cite{Bezati:2019} and the tool in~\cite{Jerbi:2012} are also based on the Orcc compiler but use Vivado HLS instead of Forge as a target. Authors in~\cite{Siret:2010} directly generated VHDL from the Orcc intermediate representation.

\begin{lstlisting}[language=C,caption=Action selection controller used in Orcc.,label={c:dummy_action_selection}]
void Filter(int param, InputStream IN, OutputStream OUT){
 if(IN.available_token(1) && pred(param, IN.peek_token(1))){
  if(OUT.available_space(1))
   t0();
 } else {
  if( IN.available_token(1) )
   t1();
 }
}
\end{lstlisting}

A key difference of this work from other dataflow-to-hardware efforts is our use of the AM 
model to select actions for execution. Other \Cal tools use a “basic” controller model that checks each consecutive execution of an actor against all of its firing conditions to select the action to execute. Consider the 
actor \texttt{Filter} from Listing~\ref{cal:filter}.
Listing~\ref{c:dummy_action_selection} shows how other \Cal-to-hardware tools implement action selection.
If an actor has many actions, the number of firing conditions increases, and the cost of action selection increases. As a consequence, the latency of the actor increases too. For example, consider when the \texttt{Filter} actor has an input token in its
input port, but no space to write the output, and the \texttt{pred} function is true. 
The controller on Listing~\ref{c:dummy_action_selection} will always check first if there is input and if the result of \texttt{pred} is true.
If so, it fails because there is no output space and exits, thus, losing a cycle (or more depending on \texttt{pred}'s latency) by retesting these conditions when the controller is next started. By comparison, the AM controller remembers the conditions that were checked and so can execute action \texttt{t1} when output space is available.

SysteMoC is an actor language~\cite{systemoc:19} based on SystemC that facilitates classification of 
models of computation for hardware-software co-synthesis. SystemCoDesigner~\cite{systemcodesigner:09} is a
fully automated hardware-software design space exploration framework for SysteMoC programs on single-core SoCs with an embedded FPGA. Our work targets a broader
range of platforms, from heterogeneous, multi-core SoCs to datacenter FPGA-accelerated platforms.

TornadoVM~\cite{Fumero:19} is a virtual machine for managing Java-based streaming tasks on heterogeneous platforms incorporating 
CPUs, GPUs, and FPGAs. It can dynamically reconfigure tasks written in Java on available hardware resources. It precompiles each task to
an FPGA bistream and then dynamically reconfigures programs on different hardware based on run-time information and program input size. We perform a static partitioning in which multiple tasks
can be offloaded to the FPGA, while their work only accelerates a single task at a time.

LINQits~\cite{LINQits:13} takes a program written in LINQ and accelerates query operation with a template library of hardware accelerators for an embedded heterogeneous MPSoC.
User-defined anonymous functions are plugged into the template libraries through HLS  synthesis. Unlike our work, LINQits focuses on accelerating the whole program (or kernel) while StreamBlocks provides
a flexible boundary for acceleration.

SCORE~\cite{SCORE:06} pioneers time-domain multiplexing to fit applications that require more resources than 
what is available on FPGAs and uses loadtime and runtime techniques for 
scheduling and placement. However, SCORE does not envision multi-threaded heterogeneous execution.

Finally, TURNUS~\cite{Casale:2013:SPIC} is a design space exploration and optimization tool that uses a different methodology for partitioning a dataflow program. TURNUS uses post-mortem execution traces of a dataflow application and profiling information to find partitions for homogeneous platforms. StreamBlocks supports both homogeneous and heterogeneous partitioning.


\section{Evaluating Partitioning}
\label{sec:evaluation}

\begin{table*}[t]
\centering
    \begin{tabular}{ l*7cc}
                                    &                       &                  & \multicolumn{5}{c}{\textbf{Throughput}}                                                                                             &                                  \\
         \textbf{Benchmark}         & \textbf{Actors}       &  \textbf{FIFOs}  &  \textbf{hardware}   & \textbf{single}    & \textbf{many}                   & \textbf{unit}                    &   \textbf{speedup} &   \textbf{Domain}                 \\
         \toprule
         \textbf{JPEG Blur}         &    104                &  210             &    {881.96}          &  {161.33}          &    {127.06}                     & {frames/second}                  &    {5.47}          & Decoding/format conversion/stencil \\
         \textbf{RVC-MPEG4SP}       &    60                 &  123             &    {1858.62}         &  {868.61}          &    {472.44}                     & {frames/second}                  &    {2.14}          & Video Decoding                     \\
         \textbf{Smith-Waterman}    &     8                 &   30             &    {12911.56}        &  {3967.07}         &    {204.39}                     & {alignments/second}              &    {3.25}          & Sequence alignment                 \\
         \textbf{SHA1}              &    20                 &   26             &    {130.64}          &  {53.75}           &    {177.66}                     & {MiB/second }                    &    {0.74}          & Encryption                         \\
         \textbf{Bitonic Sort}      &     28                &   57             &    {6443K}           &  {5215K}           &    {5477K}                      & {sort/second}                    &    {1.23}          & Hardware sorting                   \\
         \textbf{FIR}               &     34                &    45            &    {56}              &  {7.2}             &    {0.16}                       & {MiB/second}                     &    {7.8}           & 1D convolution                     \\
         \textbf{IDCT}              &     7                 &   9              &    {1612K}           &  {979K}            &    {2039K}                      & {macroblock/second}              &    {1.64}          & Inverse cosine transform                   \\
         \bottomrule
    \end{tabular}
    \caption{An overview of the benchmarks on Intel Xeon 6234 + Alveo U250.}
    \label{tab:benchmarks}
\end{table*}

We performed our experiments on two systems.
The first was a single node in the ETH Zurich XACC cluster that contains an Intel Xeon Gold 6234 8-core (16-thread) 3.3GHz processor and an Alveo U250 accelerator
card connected to the system through a Gen3 PCIe x16 slot. This system is representative of data center accelerator platforms, with a high-end processor and a large FPGA. 
To evaluate our approach for embedded systems, we used the ZCU106 development board containing an MPSoC with 4 ARM64 cores running at 1.2GHz and a medium-size FPGA.

\subsection{Benchmarks}
We used a suite of benchmarks of varying complexity from different application domains. 
Table~\ref{tab:benchmarks} summarizes these benchmarks, with performance on the U250 platform. 
The benchmarks differ widely:
\begin{itemize}
    \item \textbf{JPEG Blur} performs 8 coarse tasks: parsing, Huffman decoding, dequantization, 
        cosine transform, macro block to raster conversion, Gaussian blur filter, and raster to macro block conversion. 
        The Gaussian blur kernel implementation is based on the work of Cong et al~\cite{Cong_stencil:14}.
    \item \textbf{RVC-MPEG4SP} video decoder is a \textit{reference} implementation of the RVC-MPEG4SP ISO/IEC 14496-2 MPEG-4 standard. This 
    design has been used in many \Cal-related publications. The decoder is composed of 7 sub-networks: parser, 3 texture decoding networks, and
    3 motion compensation networks.
    \item \textbf{Smith-Waterman} is an implementation of the Smith-Waterman string matching algorithm for DNA alignment~\cite{Casale:2017}.
    \item \textbf{SHA1} is a straightforward implementation of the SHA1 algorithm with eight SHA1 compute engines. Each engine has 
    a padding actor and a compute actor.
    \item \textbf{Bitonic sort} is an eight-element bitonic sort implementation.
    \item \textbf{FIR} a 64-tap pipelined FIR filter.
    \item \textbf{IDCT} Inverse cosine discrete transformation used in video and image decoding.
\end{itemize}

The throughput numbers in Table~\ref{tab:benchmarks} reflect three scenarios:
\begin{itemize}
    \item \textbf{hardware.} All actors are placed on the FPGA with the exception of 2 or 3 actors that perform
    file and IO operations.
    \item \textbf{single.} All actors are placed on a single software thread.
    \item \textbf{many.} Each actor runs on its own thread, e.g., if there are 104 actors in a system, then 104 threads are spawned.
\end{itemize}

As evident in Table~\ref{tab:benchmarks}, using a thread per actor frequently degrades performance
because of the cost of thread scheduling and inter-thread communication. Furthermore, placing all actors in hardware does not
necessarily result in the best performance.
The three throughput measurements in Table~\ref{tab:benchmarks} represent the three corners of the design space, but not necessarily the best performing points. 
In fact, we demonstrate that when multiple software threads work in tandem with the FPGA better performance is achieved.

Below, we focus only on JPEG Blur and RVC-MPEG4SP,
as they are fairly large and contain a set of computational kernels with dynamic behavior.

\subsection{Design Space Exploration}\label{sec:dse}
In this section, we use
the MILP formulation presented in Section~\ref{sec:hetero:partitioning} to automatically explore the design space of the JPEG Blur
and RVC-MPEG4SP benchmarks on both platforms.
To do so, we solve the MILP formulation for a fixed number of threads, with and without an FPGA.
We vary the number of threads from 2-8 on the datacenter platform and 2-4 on the embedded platform
the Xeon processor has 8 cores and the ARM processor has 4. This produces many
multi-threaded software and multi-threaded heterogeneous partitions. To effectively utilize OpenCL read and write bandwidth,
in the heterogeneous solutions we allocate 1-4 MiB OpenCL buffers.
Inside the FPGA, FIFO sizes are 
set by the values in the \Cal code or set to 4096 on Alveo U250 and 512 on ZCU106 (default value) if unspecified. 
Likewise, for multi-threaded software,
we do not override the buffer configuration since buffers that do not span software and hardware are not as sensitive to payload size. The selection of buffer size is reflected in the 
in the MILP formulation.
Without this design choice, hardware performance is severely bottlenecked by poor hardware-software data transfers and, in fact, the MILP optimizer fails to find meaningful hardware partitions.

\begin{figure*}[h]
    \centering
    \subfloat[]{
        \includegraphics[width=0.5\linewidth]{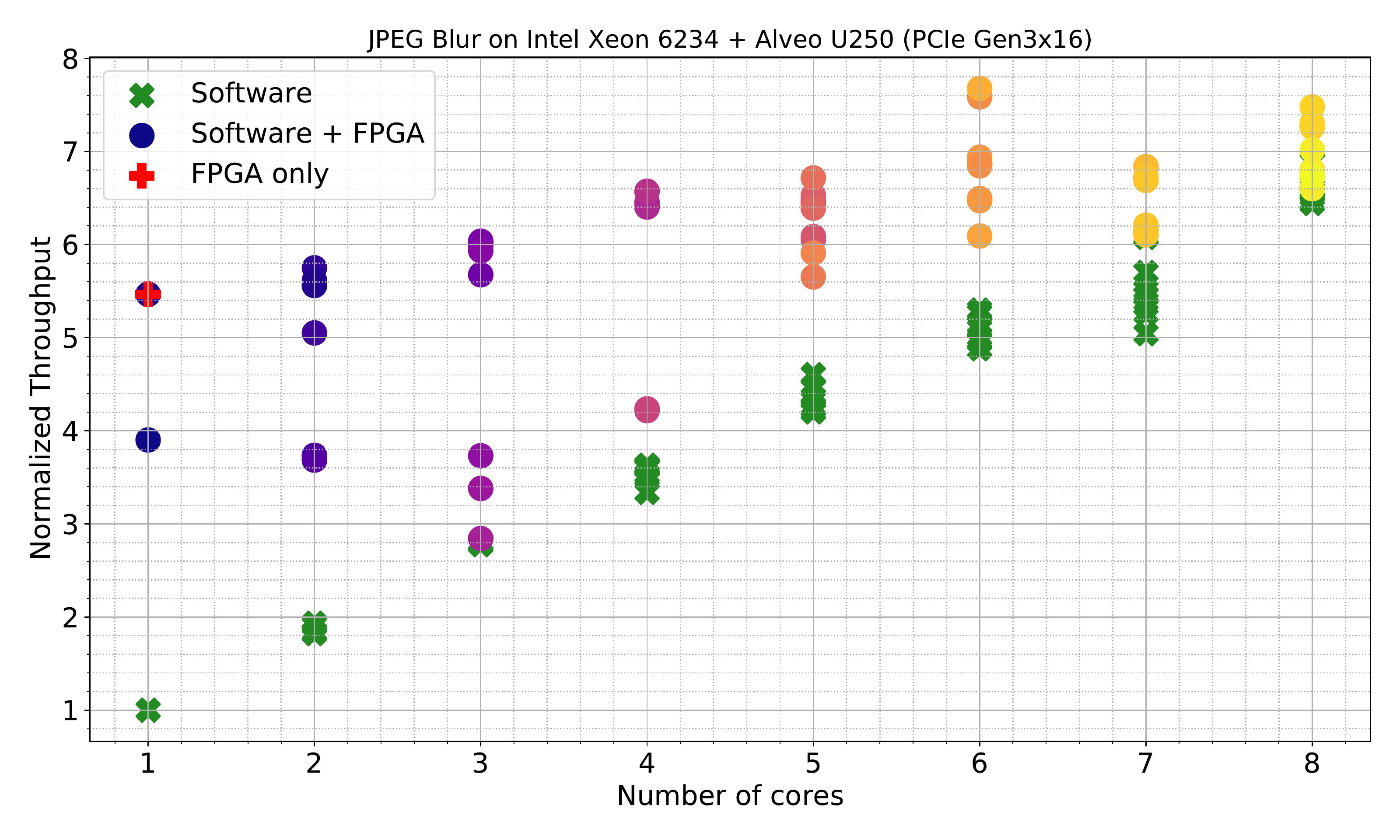}
        \label{fig:jpegblur_pp}
    }
    \subfloat[]{
        \includegraphics[width=0.5\linewidth]{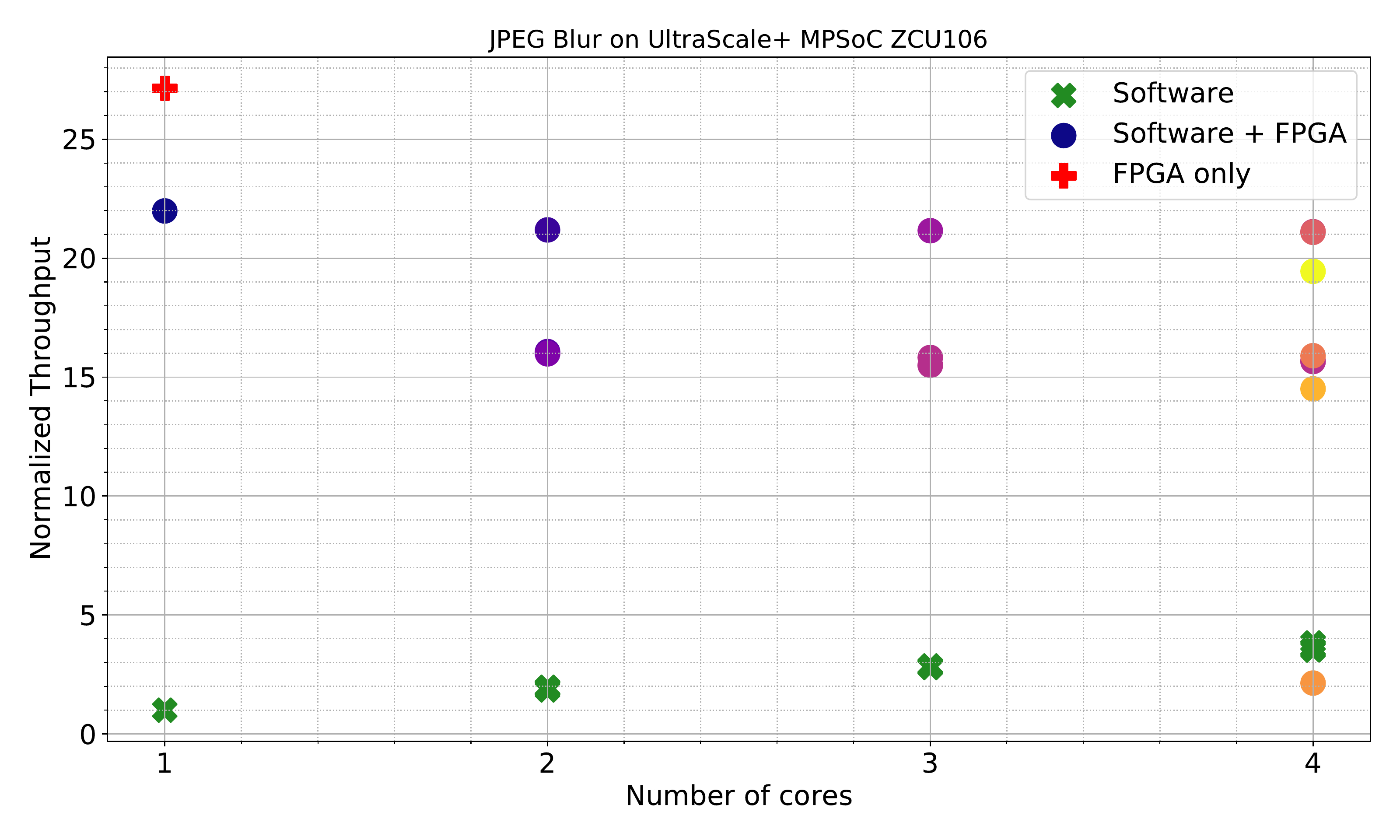}
        \label{fig:jpegblur_pp_zcu}
    }
    \caption{Evaluated JPEG Blur design points on (a) Alveo U250 and (b) ZCU106 platforms, each hardware partition 
    has a unique color.}
    \label{fig:jpegblur_perf}
\end{figure*}

The MILP optimizer takes four inputs: (i) SystemC profile with per-actor clock cycle count obtained through
co-simulation, (ii) software profile with per-actor timing information measured with CPU hardware counters, (iii) 
software FIFO bandwidth measured with CPU hardware counters, and (iv) OpenCL read and write bandwidth for a range of
buffer sizes measured with OpenCL event counters. Observe that (i) is FPGA clock cycles, (ii) and (iii) are CPU cycles, and (iv) is nanoseconds. We convert all these numbers
to nanoseconds by ``guessing" a final FPGA operating frequency and using the advertised processor clock speed. The outcome of our experiments are summarized in Table~\ref{tab:dse}. 

\begin{table}[hb]
    \centering
        \begin{tabular}{l l l  c c}
        \multicolumn{3}{c}{\textbf{Benchmark}}                                                                  &   \textbf{JPEG Blur}            &      \textbf{RVC-MPEG4SP}  \\
        \toprule
        \multirow{6}{*}{\rotatebox{90}{U250}}  &              \multicolumn{2}{l}{\textbf{baseline} (fps)}                           &   {161.33}   &   {868.61}  \\
                                               
                                               & \multirow{2}{*}{\rotatebox{00}{\textbf{software}}}     &   \textbf{partitions}     &   {63}     &   {70}  \\
                                               &                                                        &   \textbf{speedup}        &   {6.90}   &   {4.78} \\
                                               
                                               & \multirow{2}{*}{\rotatebox{00}{\textbf{software + FPGA}}}     &   \textbf{partitions}     &   {67}     &   {66}  \\
                                               &                                                        &   \textbf{bitstreams}     &   {34}     &   {38}  \\
                                               &                                                        &   \textbf{speedup}        &   {7.68}   &   {4.40} \\
        \midrule
        \multirow{6}{*}{\rotatebox{90}{ZCU106}}  &              \multicolumn{2}{l}{\textbf{baseline} (fps)}                           &   {22.13}   &   {109.87}  \\
                                               
                                               & \multirow{2}{*}{\rotatebox{00}{\textbf{software}}}     &   \textbf{partitions}     &   {30}   &   {  30}  \\
                                               &                                                        &   \textbf{speedup}        &   {3.83}   &   {3.92} \\
                                               
                                               & \multirow{2}{*}{\rotatebox{00}{\textbf{software + FPGA}}}     &   \textbf{partitions}     &   {17}   &   {   14}  \\
                                               &                                                        &   \textbf{bitstreams}     &   {12}   &   {    8}  \\
                                               &                                                        &   \textbf{speedup}        &   {27.14}   &   {14.44} \\

        \bottomrule
    \end{tabular}
    \caption{Design space exploration summary of the JPEG Blur and RVC-MPEG4SP benchmarks on 
    the U250 data-center and the ZCU106 embedded platforms. 
    Speedup numbers correspond to the maximum performance achieved with and without hardware. Bitstreams
    counts the number of unique hardware partitions (which may have multiple software-thread partitions).}
    \label{tab:dse}
\end{table}

\subsubsection{JPEG Blur}

Fig.~\ref{fig:jpegblur_perf} illustrates the various evaluated design points of the JPEGBlur 
benchmark on the Alveo U250 and ZCU106 platforms. 
On Alveo U250, there are 34 hardware partitions (i.e., one or more actors on hardware -- bitstreams in Table~\ref{tab:dse}).
Since the remaining software actors can be further partitioned across threads, 
there is a total of 67 heterogeneous partitions. Furthermore, we found 63 multi-threaded partitions that do not use hardware at all.
On ZCU106, design space exploration yielded 30 software-only partitions and 12 hardware partitions resulting in 17 heterogeneous partitions.

The heterogeneous points in Fig.~\ref{fig:jpegblur_perf} are colored based on the hardware partition used at each point. The throughput is also normalized to the baselines reported in 
Table~\ref{tab:dse}.

Evidently, JPEG Blur can scale almost linearly with number of cores without hardware acceleration on both
platforms. When an FPGA is used, scaling is not as well-behaved.
We see a sub-linear speedup on Alveo U250 (Fig.~\ref{fig:jpegblur_pp}) where heterogeneous 
performance is better, in general.
Conversely, heterogeneous performance slightly 
degrades with additional cores on the embedded ZCU106 platform (Fig.~\ref{fig:jpegblur_pp_zcu}).
The best performance is the hardware-only design.
Note that all of these alternatives were explored \emph{without} code-rewriting or refactoring.

\begin{figure}[h]
    \centering
    \includegraphics[width=\linewidth]{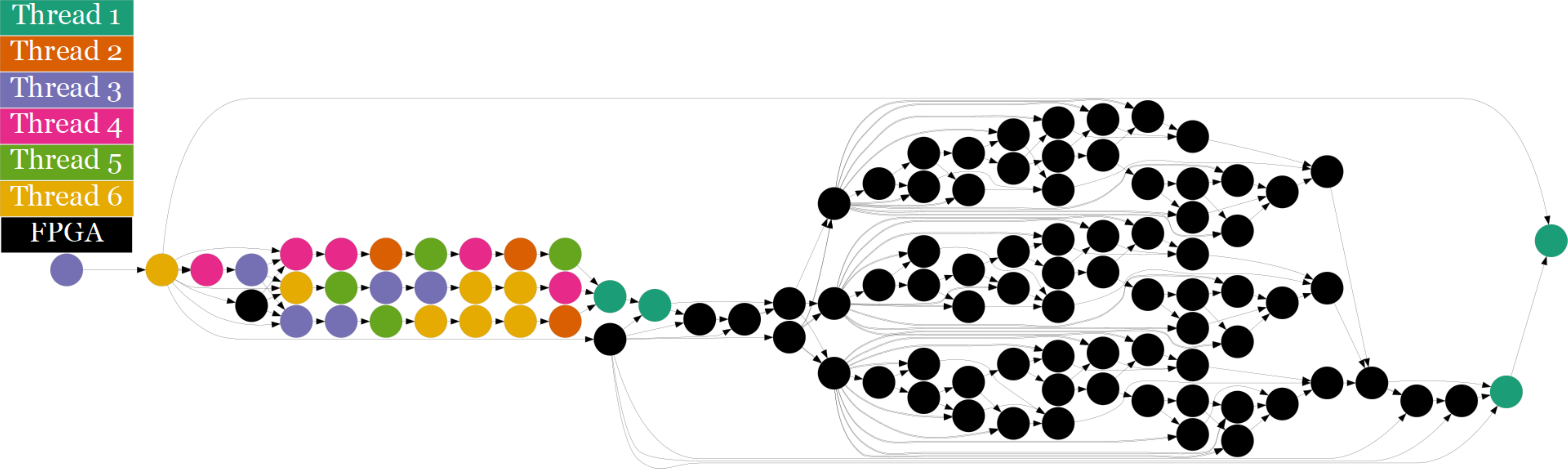}
    \caption{JPEG Blur configuration with the highest throughput.}
    \label{fig:jpegblur_pp_best}
\end{figure}

Fig.~\ref{fig:jpegblur_pp} demonstrates our claim that the optimum design point is not necessarily 
a hardware-only or software-only configuration, but rather a mixture. Without a 
common programming model to facilitate exploration of many partitions, exploring these design points
would be a tedious process of code rewriting.
Note, however, that due to the heuristic partitioning method, 
the discovered design points are not necessarily the global optimum either. 
Our claim is reinforced by Fig.~\ref{fig:jpegblur_pp_best}, which depicts the best performing partition (on Alveo U250). 
This partition places a substantial number of
actors on 6 software threads and boosts the speedup from about 5.5x (all hardware) to 7.7x. This non-trivial
partition can be easily overlooked when partitioning takes place early in the implementation.

\subsubsection{RVC-MPEG4SP}
\begin{figure*}[t]
    \centering
    \subfloat[]{
        \includegraphics[width=0.5\linewidth]{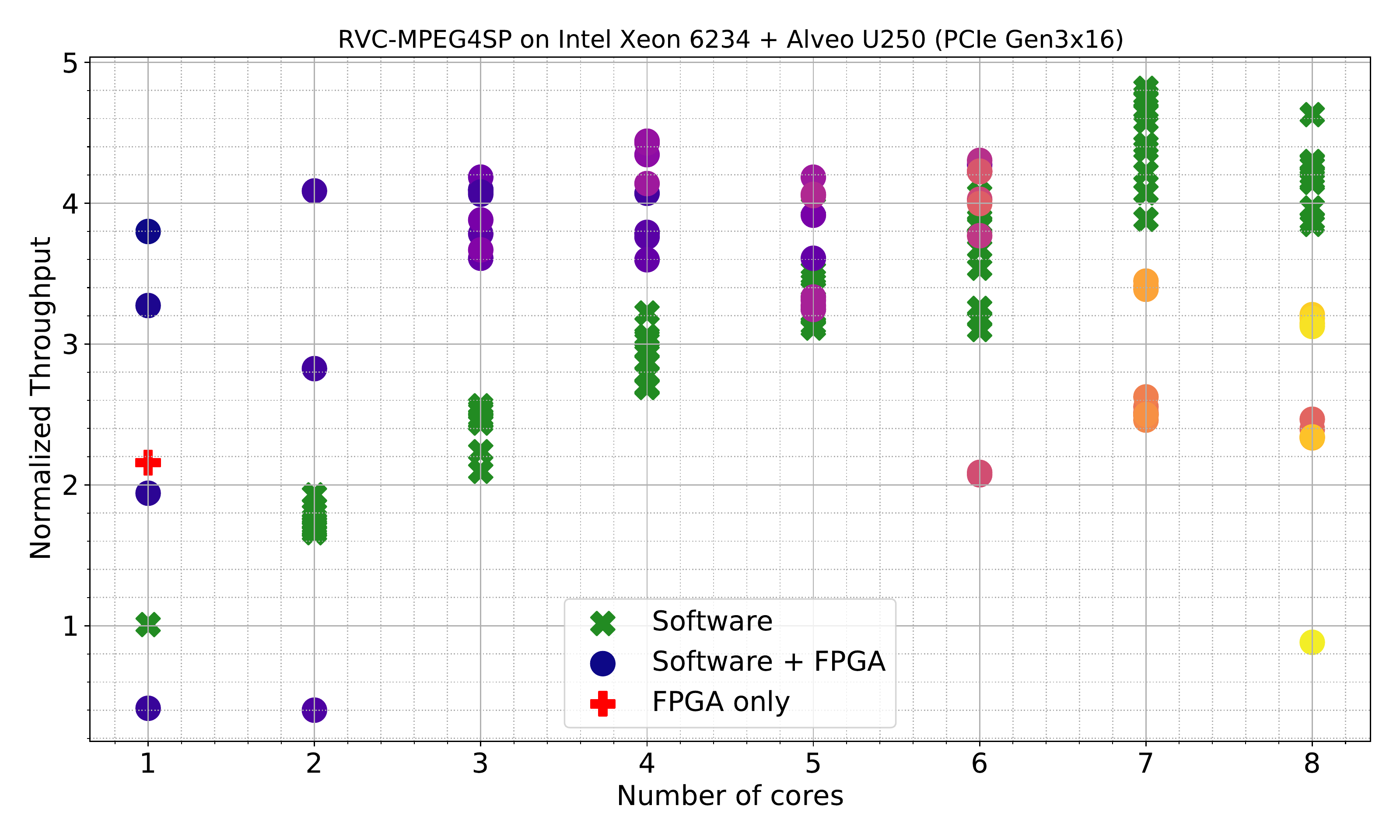}
        \label{fig:rvc_u250}
    }
    \subfloat[]{
        \includegraphics[width=0.5\linewidth]{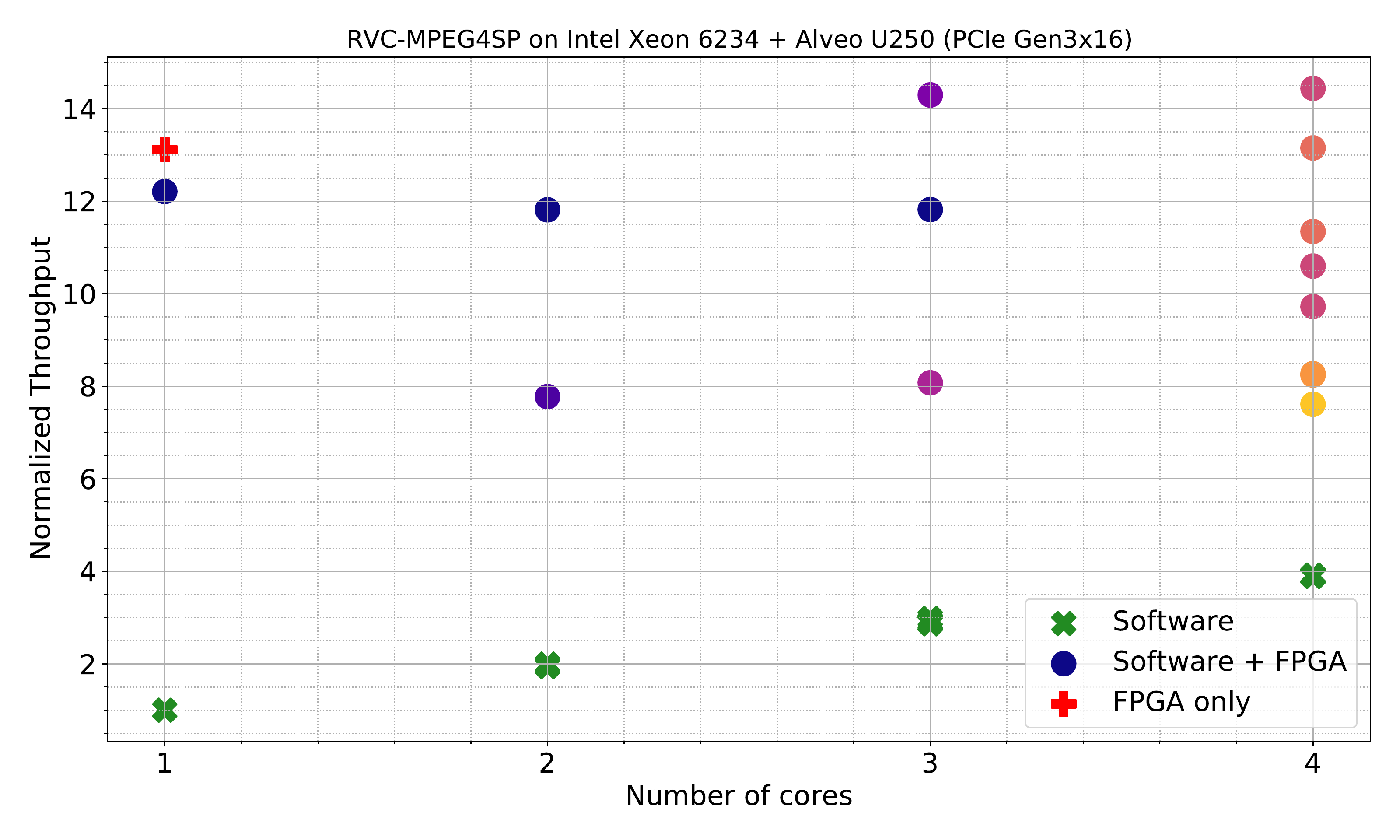}
        \label{fig:rvc_zcu}
    }
    \caption{Evaluated RVC-MPEG4SP design points on (a) Alveo U250 and (b) ZCU106 platforms, each hardware partition 
    has a unique color.}
    \label{fig:rvc_perf}
\end{figure*}

We discovered 66 heterogeneous partitions on Alveo U250, among which 38 correspond to a unique subset of actors on hardware.
Without the FPGA, software partitioning yielded 70 configurations with 2-8 threads.
With the ZCU106 platform, we found 30 software partitions and 14 heterogeneous partitions based on 8 unique hardware partitions.
This experiment is summarized in Fig.~\ref{fig:rvc_perf}, where the throughput is normalized to the single-thread 
execution in Table~\ref{tab:dse}. 

RVC-MPEG4SP scales differently than the JPEG Blur benchmark on Alveo U250. Software enjoys linear acceleration 
up to 7 cores but then the performance drops, marking the limit of scalability. 
Similar to software scaling, the sub-linear heterogeneous scaling experiences a down-turn at 4 cores and 
eventually software performance surpasses heterogeneous performance.

On the ZCU106 platform, software scaling is almost perfectly linear for the 4 available cores.
The heterogeneous performance sees a marginal increase with additional cores, unlike Fig.~\ref{fig:jpegblur_pp_zcu}, where 
heterogeneous performance experiences a slight dip with more cores.

We found a single-core heterogeneous partition that is 1.5x faster than an all-hardware solution, a point in the design space that could be easily overlooked.

\begin{figure}
    \centering
    \includegraphics[width=\linewidth]{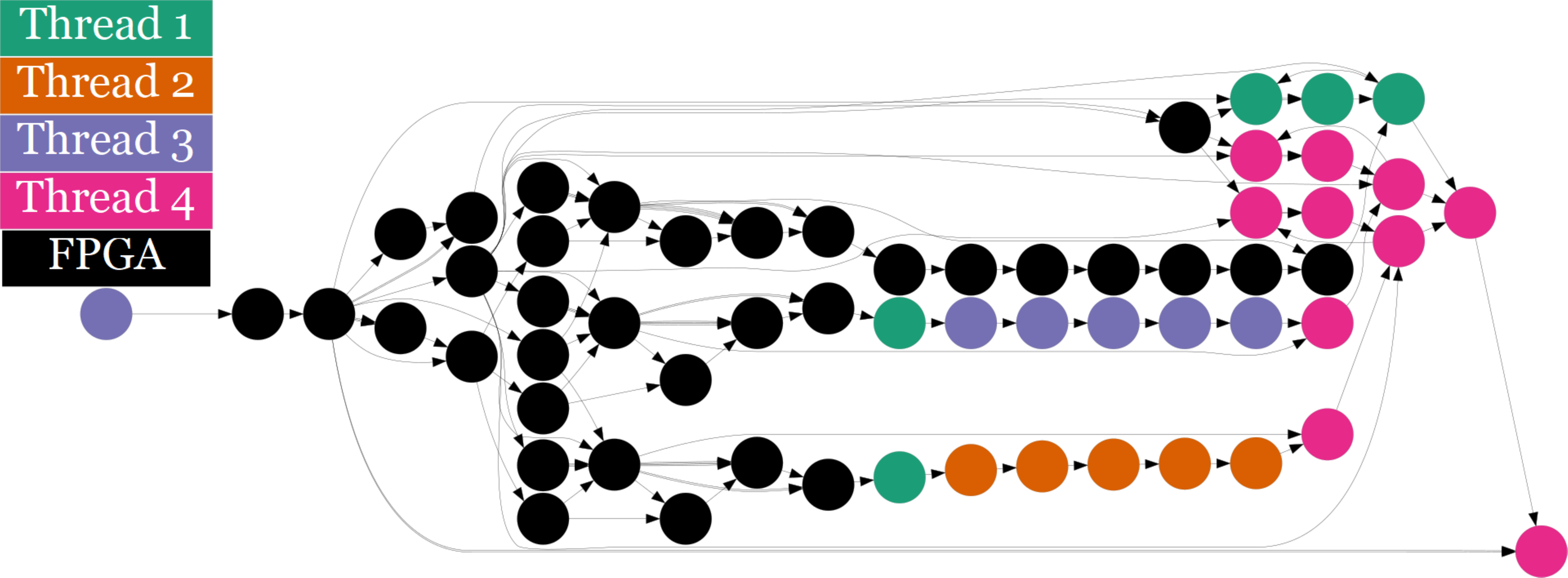}
    \caption{Best performing heterogeneous partition for the RVC-MPEG4SP benchmark.}
    \label{fig:rvc_u250_best}
\end{figure}

The best heterogeneous Alveo U250 partition is outlined in Fig.~\ref{fig:rvc_u250_best}. Such a valuable 
configuration may be simply left out in a manual development process.

\subsection{Multi-Objective Optimization}

We showed that given some cores, our methodology can find meaningful heterogeneous partitions that perform better. Of course, with performance and resource objectives, a developer seeks a point in the design space that \emph{just about satisfies} his or her objectives. For instance, if the target performance is 
4.8x the baseline, then a software-only partition should be used for the RVC-MPEG4SP on Alveo U250. Alternatively,
if a 4.0x acceleration is \emph{enough}, one would naturally choose the highest point with 2 cores on Fig.~\ref{fig:rvc_u250}.
The design objective can be many-fold, namely, one can limit the number of cores or the resources used on the FPGA.

Multi-objective optimization is a future direction of our work. The constraint-based formulation 
that we used to solve the partitioning problem can be extended to handle other objectives as well. A common 
approach is to define the optimization metric as a linear combination of
individual objectives. Concretely, if it is desired to minimize execution time and FPGA resource usage 
simultaneously, we can minimize $T + \alpha R$, where $T$ is execution time, $R$ is the resource
utilization and $\alpha$ is a scaling factor that assigns a relative weight to the objectives.

\subsection{Exploration Time}

There are three stages in our design space exploration methodology: (i) profiling, (ii) solving the MILP formulation, 
and (iii) evaluating through synthesis and implementation. 
Not surprisingly, most of the time is spent in the last step.

With the current FPGA tools, synthesis and implementation time takes up to several hours. In our experiments with the U250 board
and Vitis 2020.1 tool chain, each heterogeneous point in the Fig.~\ref{fig:jpegblur_pp} and~\ref{fig:rvc_u250} required 3-5 hours to compile---essentially a week of compilation for each benchmark.

Fortunately, all of points discovered by the MILP formulation can be compiled in parallel. In this evaluation,
we distributed  compilation to three 24-core (48-thread) servers with 256 GiB of RAM,
which enabled us to compile 20 points concurrently (before running out of memory), essentially turning a week into a single day.

Hardware profiling time is highly dependent on the actor network and the input size.
In both benchmarks, we used down-sampled inputs to keep the profiling time below 1 hour. Software profiling, on the other hand, takes only seconds. 

We use a fast industrial MILP solver and constrained the run time to be less than 5 minutes 
per core configuration (e.g., 40 minutes with 8 cores). 
We plan to evaluate an open-source solver back-end as well.

\section{Conclusion}
\label{sec:conclusion}

This work presents a new dataflow compiler for FPGA-based heterogeneous platforms. With the \Cal programming language and its
efficient translation to the actor machine model and with our compiler, it is possible to express and compile complex streaming programs for both software and hardware targets. Seamless hardware-software execution of streaming functions written in a single source language
opens up new avenues in the design space exploration of a heterogeneous application. In addition, automated partitioning
significantly reduces the manual effort required to rewrite and optimize complex programs.

We described how the StreamBlocks compiler generates and interconnects the computation on software and hardware heterogeneous platforms. 
Moreover, the StreamBlocks partitioning tool uses profiling information to 
efficiently map the computations onto both software or hardware. 
Our experiments demonstrate this approach's potential by exploring multiple partitionings of two large programs.

Our future work includes 1) platform-specific code optimization such as automated HLS directive injection and 
post-partitioning optimization and 2) multi-objective design space exploration.

\section{Appendix}\label{sec:appendix}

\begin{figure*}[h]
  \centering
  \subfloat[]{
      \includegraphics[width=0.5\linewidth]{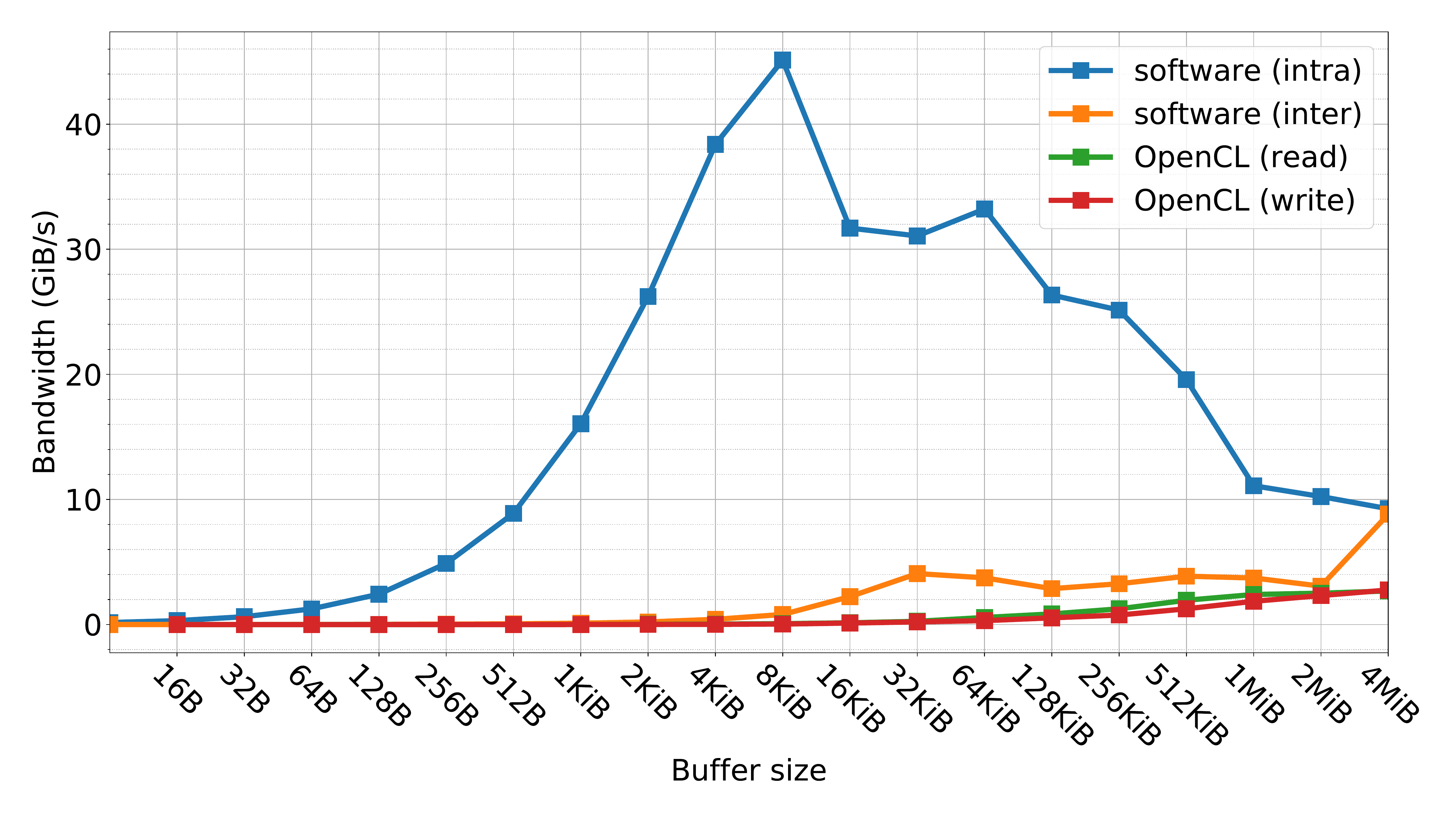}
      \label{fig:fifo_cost_u250}
  }
  \subfloat[]{
      \includegraphics[width=0.5\linewidth]{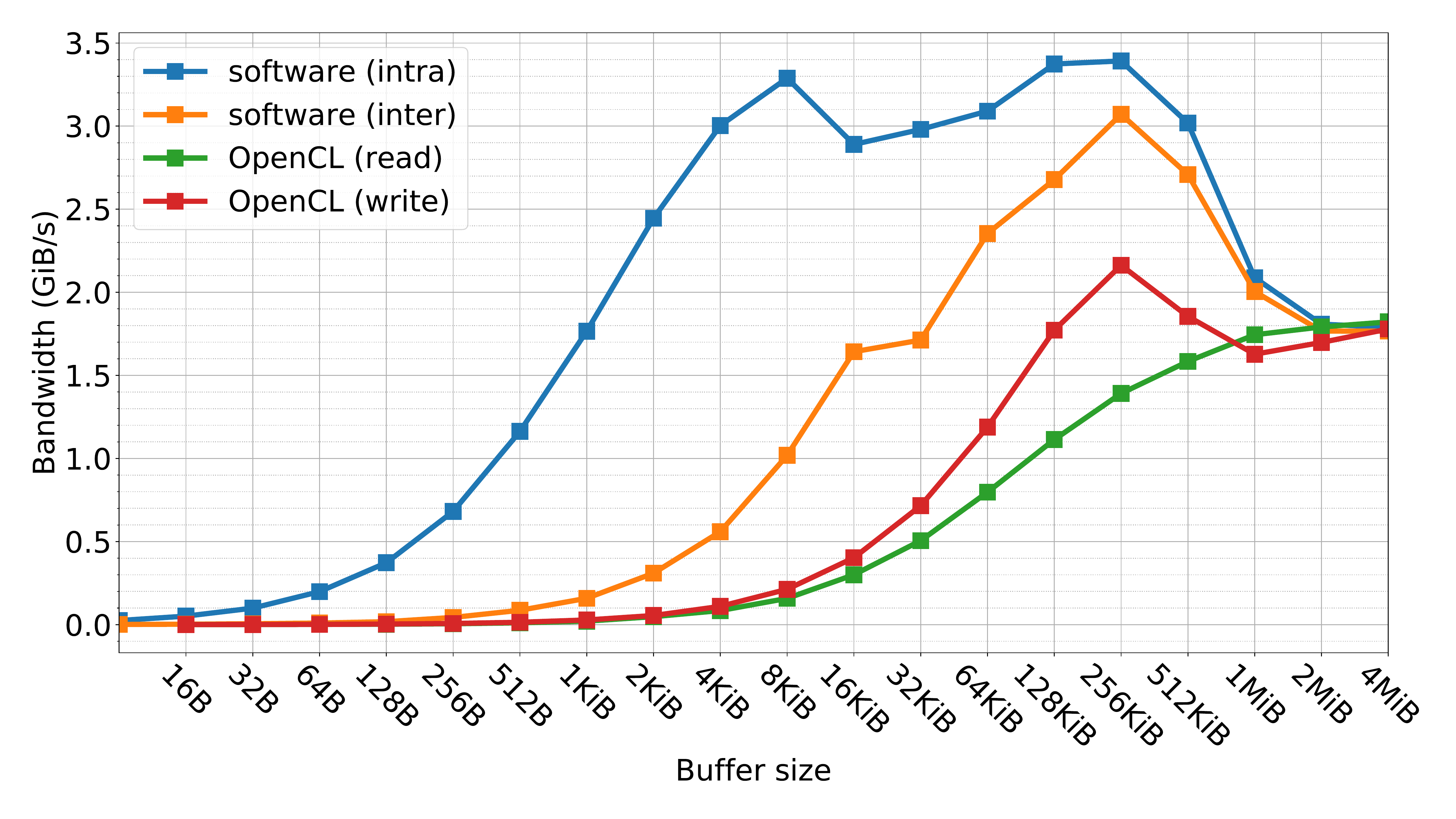}
      \label{fig:fifo_cost_zcu}
  }
  \caption{Measured FIFO communication bandwidth on (a) Alveo U250 and (b) ZCU106.}
  \label{fig:fifo_cost}
\end{figure*}

\subsection{MILP Formulation Details}
$T_{plink}^{write}$ and $T_{plink}^{read}$ are upper bound estimates of
the runtime cost of transferring data between host and FPGA. 
We represent the set of all connection as the set $C$ and such that $(s, t) \in C$
denotes a connection from source port $s$ to target port $t$. Furthermore, 
$s.a$ and $t.a$ denote the actor that the port belongs to.

We obtain the number of tokens traversed a connection $(s, t)$ through 
profiling and denote it as $n_{(s, t)}$. Every connection also
has an associated buffer size $b_{(s, t)}$. 
OpenCL read and write operations are most efficient if a full buffer 
(i.e., $b_{(s, t)}$ tokens) are transferred.

We measured the OpenCL transfer operation times from queueing to completion 
using OpenCL event counters to obtain a function $\xi_r(b)$ and $\xi_w(b)$ 
that models OpenCL read and write time given a 
buffer configuration $b$.
Using this function we can estimate the best case time required to write $n$
tokens over buffers with capacity $b$ as: 

\begin{equation}
  \tau_w(n, b) = 
  \begin{cases}
    \xi_w(n) & n \leq b \\
    \xi_w(b) \times \lfloor \frac{n}{b} \rfloor + \xi_w(n\bmod{b}) & n > b
  \end{cases}
\end{equation}

Here, $ \lfloor . \rfloor$ and $mod$ are the floor and modulo operators. To 
estimate read times, a function $\tau_r(n, b)$ is similarly defined by replacing 
$\xi_w(b)$ with $\xi_r(b)$. Now we can estimate PLink read and write times as follows:

\begin{equation}
  \begin{split}
    T_{plink}^{write} = & \sum_{(s, t) \in C} (\neg d^{s.a}_{accel} \land d^{t.a}_{accel}) 
                              \tau_w(n_{(s, t)}, b_{(s, t)}) \\
    T_{plink}^{read} = & \sum_{(s, t) \in C} (d^{s.a}_{accel} \land \neg d^{t.a}_{accel})
                              \tau_r(n_{(s, t)}, b_{(s, t)})                 
  \end{split}
\end{equation}

Where $\land$ and $\neg$ are the logical \emph{conjunction} and \emph{negation} operators 
respectively. The conjunction expressions ensure that read and write times are 
only accounted for connections with one port on the FPGA and the other on the CPU.

Similarly, for every partition $p \in P_{thread}$, we define
\begin{equation}
    t^p_{intra} = \sum_{(s, t) \in C} (d^{s.a}_p \land d^{t.a}_p) \tau_{intra}(n_{(s, t)}, b_{(s, t)})
\end{equation}
Where $\tau_{intra}$ is models the intra-core communication time and is obtained
by profiling software FIFO read and write bandwidth. To do this, we 
measure the roundtrip times of sending a token from one port and receiving in 
from another port going through a pass-through actor. Therefore, we use the 
same bandwidth value for both read and write 
(i.e., time for read \emph{or} write is round-trip time divided by 2).

However, notice that if a connection $(s, t)$ has its source actor on the 
first partition (i.e., $p_1$ which also contains the PLink) and its target 
actor is on the FPGA, the data should first be copied to the PLink and then 
to the device. This can be modeled by $t^{plink}_{intra}$ as follows:

\begin{equation}
  \begin{split}
    t^{plink}_{intra} & = \\
    \sum_{(s, t) \in C}  \left( (d^{s.a}_{p_1} \land d^{t.a}_{accel})  \lor
      (d^{s.a}_{accel} \right. & \left.  \land d^{t.a}_{p_1}) \right) 
       \tau_{intra}(n_{(s, t)}, b_{(s, t)})  
  \end{split}
\end{equation}

Where $\lor$ is the logical disjunction operator. 
We can now define the local core communication cost for each partition $p \in P_{thread}$
as:

\begin{equation}
  T^p_{intra} = 
  \begin{cases}
    t^p_{intra} & p \in P_{thread} \setminus \{p_1\} \\
    t^p_{intra} + t^{plink}_{intra} & p = p_1
  \end{cases}
\end{equation}

Since all the threads operate in parallel, then the total intra-core communication
time is obtained as the maximum of individual ones:

\begin{equation}
  T_{intra} = max(\{T_{intra}^p : p \in P_{thread}\})
\end{equation}

Finally, we estimate the core to core communication cost as follows:

\begin{multline}
  T_{inter} = 
  \sum_{(s, t) \in  C} 
    \Biggl( \Biggr. \\
      \sum_{q \in P_{thread} \setminus \{p_1\}}
        \left( 
          \left(
            (d^{s.a}_{p_1} \lor d^{s.a}_{accel}) \land d_q^{t.a}
          \right)
          \lor  \right. \\ \left.
          \left(
            d_q^{s.a} \land (d^{t.a}_{p_1} \lor d^{t.a}_{accel})
          \right)
        \right) \\
        + \sum_{p \in P_{thread} \setminus \{p_1\}}\sum_{q \in P_{thread} \setminus \{p, p_1\}} 
        \left(
          \left(d^{s.a}_{p} \land d^{t.a}_{q}\right) 
          \lor
          \left(d^{s.a}_{q} \land d^{t.a}_{p}\right)
        \right) \\
    \Biggl. \Biggr) \tau_{inter}(n_{(s, t)}, b_{(s, t)})
\end{multline}
This formula includes a cost for all of the connection that cross a thread. Notice 
how the first partition $p_1$ which contains the PLink is treated slightly differently
to handle connections from a thread $q \neq p_1$ to $accel$ or $p_1$ and vice versa.

With the derivation of $T_{inter}$, $T_{intra}$ and $T_{plink}^{read}$ and $T_{plink}^{write}$ 
our formulation is complete.

\subsection{MILP Model Accuracy}

The MILP formulation is a crude and optimistic model for performance. It tries 
to model a data-dependent and dynamic system with a series of \emph{linear} equations.
To better understand the quality of our modelling, we first 
describe its sources of inaccuracy.
\begin{enumerate}
    \item \textbf{Assuming perfect dataflow} The hardware and multi-threaded 
    execution times (excluding communication cost) is modeled as the maximum of 
    individual hardware actor and software thread execution times respectively
    (see Section~\ref{sec:hetero:partitioning}).
    This oversimplified model then inherently \emph{disregards} any data-dependent behavior 
    and assumes a perfect task-pipelined execution on hardware and no data-starvation on software.
    \item \textbf{Assuming perfect transfers}
    The communication cost model included in the appendix assumes that 
    the FIFOs in software and those that cross the PCIe bus operate at full efficiency.
    This means our model does not capture executions in which data-dependence leads to poor 
    communication efficiency.
    In fact, in Fig.~\ref{fig:rvc_u250} there are 3 heterogeneous points that perform poorly. 
    In the single-core heterogeneous point at the lower left of 
    the Fig.~\ref{fig:rvc_u250}, on average 25.12 OpenCL kernel calls are made to 
    decode 10 frames while for the best heterogeneous point (which uses 4 threads) 
    this number is 1.9 OpenCL kernel calls. 
    We have also measured the runtime bandwidth of the
    output ports of the poorly performing configuration using OpenCL event counters and 
    observed that the PCIe bus was underutilized. Notice that such discrepancies between 
    our model leads to design points with undesirable quality.
    
    \item \textbf{Profiling errors}
    The $exec(\_, \_)$ and $\xi(\_)$ functions in the formulation come from 
    profiling which entails some inherent error from measurement.
    \item \textbf{Clock period approximation}
    When solving the MILP, we should "guess" a final operating 
    frequency for the FPGA, and assume a "fixed" clock speed for the host. 
    The compiled FPGA bitsream may end up operating at a clock speed different from the 
    guessed value and the fixed clock speed can not capture the effects of frequency 
    scaling present on the host.
\end{enumerate}

Considering all of the possible sources of errors described, 
the end-to-end median execution time prediction accuracy for RVC-MPEG4SP is 22.66\% 
in Fig.~\ref{fig:rvc_u250} and 20.294\% in Fig.~\ref{fig:rvc_zcu}. 
The JPEG Blur benchmarks has a 12.84\% median error 
in Fig.~\ref{fig:jpegblur_pp} on Alveo and a 34.03\% median error in Fig.~\ref{fig:jpegblur_pp_zcu} on ZCU106.

Simulation can model data-dependent behavior and therefore will give better 
performance prediction than ``static formulation" but simulating a single design point can take hours. An 
effective design space exploration would require exploring not one, but maybe tens to 
hundreds of points which makes simulation unattractive.

\subsection{Communication Cost Measurements}
In the MILP formulation, the communication cost is parameterized as a
function $\xi(b)$ that represents the measured read or write times. If 
we plot the communication bandwidth as $b / \xi(b)$, then we obtain the 
Fig.~\ref{fig:fifo_cost_u250} and Fig.~\ref{fig:fifo_cost_zcu} for 
the Alveo U250 and ZCU106 platforms respectively.

Notice the large difference 
between inter- and intra-core communication bandwidth in Fig~\ref{fig:fifo_cost_u250}. 
This is because when the two ends of 
a FIFO are on the same core (i.e., pinned thread), then the communication 
goes through the private caches (e.g., L1 and L2) without coherence traffic.
When the two ends of a FIFO are on two different cores, the tokens travel at least 
to the shared last-level cache (i.e., L3), which incurs a coherence cost.

As expected, the intra-core bandwidth increases with larger buffer transfers 
But there is an inflection point at 4KiB (the L1 data size on x86), and at 8KiB 
the bandwidth starts to descend.

Furthermore, you can observe that OpenCL read and write operations are considerably slower than software 
FIFO operations and they only partially catch up around 1MiB.

The communication cost terms in the MILP formulation penalize using too many threads or placing everything
in hardware without accounting for the considerable overhead of OpenCL read and write operations or inter-core
communication.

\section*{Acknowledgments}
We would like to the thank the Xilinx Adaptive Compute Clusters (XACC) university program for providing access to the ETHZ clusters to do our experimental results. J.W. Janneck is funded by the ELLIIT program of the Swedish government.

\bibliographystyle{IEEEbib}
\bibliography{bibfile}

\vspace{-0.8cm}

\end{document}